\documentclass[fleqn,10pt]{wlscirep}
\usepackage{rotating}
\usepackage{subfigure}

\graphicspath{{figures/}}

\title{Epidemic risk from friendship network data: an equivalence with a non-uniform sampling of contact networks}

\author[1]{Julie Fournet}
\author[1,2,*]{Alain Barrat}

\affil[1]{Aix Marseille Universit\'e, Universit\'e de Toulon, CNRS, CPT, UMR 7332, 13288}
\affil[2]{Data Science Laboratory, ISI Foundation, Torino, Italy}
\affil[*]{alain.barrat@cpt.univ-mrs.fr}


\begin{abstract}
Contacts between individuals play an important role in determining how infectious diseases spread. Various
methods to gather data on such contacts co-exist, from surveys to wearable sensors. Comparisons of data
obtained by different methods in the same context are however scarce, in particular with respect to
their use in data-driven models of spreading processes. Here, we use a combined data set describing
contacts registered by sensors and friendship relations in the same population to address this issue in a case study.
We  investigate if the use of the friendship network is equivalent to a sampling procedure performed on the sensor
contact network with respect to the outcome of simulations of spreading processes:
such an equivalence might indeed give hints on ways to compensate for the incompleteness of contact
data deduced from surveys.
We show that this is indeed the case for these data, for a specifically designed
sampling procedure, in which respondents report their neighbors with a probability depending
on their contact time. We study the impact of this specific sampling procedure on several data sets,
discuss limitations of our approach and its possible applications in the use of data sets of
various origins in data-driven simulations of epidemic processes.
\end{abstract}

\begin{document}

\flushbottom
\maketitle
\thispagestyle{empty}

\section*{Introduction}

Contact patterns between individuals, and in particular
face-to-face interactions, play an essential role in determining how infectious diseases spread within a population. 
Information about these patterns is therefore crucial to inform epidemic models and ensure relevant predictions.
Many efforts have been made to collect data about contacts between individuals using different techniques\cite{Read:2012,Eames:2015,Barrat:2015}. 
In particular, technological advances make it now possible to collect accurate data on face-to-face 
contacts with a high spatial and temporal resolution
in various contexts, by using infrastructures based on wearable sensors 
\cite{Stehle:2011b,Vanhems:2013,Salathe:2010,pentland,toth, polgreen, lehmann,Barrat:2015}.

Such infrastructures are typically deployed in specific environments (schools, hospitals, etc...) and for limited
time windows (typically few days or weeks). Moreover, even if their use has recently become more widespread,
as costs have decreased and different groups have developed similar tools, they cannot
always nor systematically be used in any context. Other types of 
data, such as surveys, diaries or online interactions, which have long been used in epidemiological
contexts \cite{Read:2012,Eames:2015}, remain thus important sources that feed models of epidemic spread.
Data on contacts between individuals coming from different types of sources are however not equivalent, as explored
quantitatively in  recent studies \cite{Smieszek:2014,Mastrandrea:2015}. 
For instance, contact diaries lead to an underestimation
of the number of actual contacts, as short contacts are not well reported, together 
with an overestimation of the duration of contacts \cite{Smieszek:2014,Mastrandrea:2015}. 
In this respect, friendship surveys, in which participants are asked to name their friends, 
might be less affected by memory biases; however, the precise
relation between the network of friendships and the network of contacts is not straightforward, even in a given context,
as friends might meet relatively rarely while many daily encounters occur
between non-friends. For instance, Ref.  \cite{Mastrandrea:2015}
reports on a high school context, where
data was collected both on actual contacts between students of $9$ classes using wearable
sensors, and on their friendship relations through a survey. It was found 
that (i) the friendship survey suffered from low participation rate,
(ii) the longest contacts corresponded to reported friendships and most friendship relations lead to actual 
face-to-face encounters, but
(iii) many short contacts did not correspond to reported friendships, 
resulting in a friendship network with much lower density than the contact network.

Here, we investigate further questions arising from this comparison. First, we assess 
how these differences between the two networks
of interactions (reported friendships vs. measured contacts) lead to different outcomes if they are used as
substrate for possible propagation events in numerical simulations of spreading processes. Indeed, although the longest
contacts are reported in the friendship survey, the non-reported many short contacts can have an important role in propagation
processes. 
Second, we ask if using the reported friendship network is equivalent, with respect to the outcome of such simulations, to 
a specific sampling of the contact network. Various sampling procedures are known to affect
networks' measured properties in different ways, and many works have studied for instance how
average degree, degree distribution, clustering or assortativity properties depend 
on the procedure and on the sample size \cite{Granovetter:1976,Frank:1978,Achlioptas:2005,Lee:2006,Kossinets:2006,Onnela:2012,Blagus:2015}.
Fewer studies have investigated how the outcome of simulations of dynamical processes in data-driven models is affected 
if incomplete data are used \cite{Ghani:1998a,Ghani:1998b,Bobashev:2008,Genois:2015}.
As incomplete network data is in fact quite common, researchers have moreover tackled the issue of
inferring network statistics \cite{Leskovec:2006,Viger:2007,Bliss:2014,Zhang:2015} 
from incomplete data. Since many networks are the support of dynamical processes, it is also
crucial to develop methods to obtain estimates of the outcome of such processes 
in the case of sampled data \cite{Ghani:1998b,Genois:2015}.
In this perspective, understanding if the differences in outcomes between contact and friendship data may be seen as 
biases due to a sampling process might then
give hints on how to compensate for such biases and how to use the information contained
in the friendship network to obtain accurate prediction on the epidemic risk, even in the absence of data on the actual contact network,
in the spirit of Ref. \cite{Genois:2015}. 

To make progresses in these directions, we 
rely here on the case study of a publicly available data set containing 
information on contacts measured by wearable sensors as well as friendship relations in the same population,
namely high school students \cite{Mastrandrea:2015}. This is indeed to the best of our knowledge
the only available data set combining contacts and friendship data in a given population.
We first show how simulations of the spread of infectious diseases using friendship data lead to a strong underestimation of 
the epidemic risk with respect to using the contact network measured using wearable sensors, as can be expected
due to the lower participation rate and the absence of many weak links in the friendship surveys. 
We then consider several sampling procedures on the contact network and compare the outcome of simulations on the sampled networks
and on the friendship network, for different values of the spreading parameters. Simple sampling methods such as uniform random sampling
of nodes or edges do not reproduce the outcome of simulations using the friendship survey. 
We then put forward a non-uniform sampling method that favors sampling of the most
important contacts of each sampled individual, and show that the outcome of spreading processes on the resulting sampled networks is
equivalent to the one obtained when the friendship network is considered. 
Note that at this stage we do not use the friendship network to provide an estimate of the real epidemic risk in the contact network,
but rather assess how using friendship data would be similar to a bias due to a specific sampling method on the real contact network.
We finally apply the sampling method to several data sets and 
study how changing its parameters
(number of nodes sampled, density of sampled network) impacts the outcome of spreading simulations.

\section*{Results}

\subsection*{Data and methodology}

We consider data collected and made publicly available
by the SocioPatterns collaboration,
which describe two types of social interactions, namely face-to-face contacts 
and friendship relations, between the same individuals,
namely students in a high school (see the SocioPatterns website \verb+http://www.sociopatterns.org/+).
The contact network, measured using wearable sensors, has $327$ nodes and $5818$ weighted edges. Each edge $(i,j)$ with weight $W_{ij}$ 
corresponds to the fact that individuals $i$ and $j$ have been in contact for a total time $W_{ij}$ during the 
deployment, which lasted one week. The friendship network, obtained through a survey, has $N_F = 135$ nodes and $E_F= 413$ unweighted edges. 
All the nodes of the friendship network (but one) belong to the contact network. 
We also note that the population under study was structured in $9$ classes (see Methods).

While Mastrandrea et al.\cite{Mastrandrea:2015} performed a quantitative comparison of the contact and friendship networks, 
we are here interested in comparing the use of these networks within the context of models of propagation phenomena. 
To this aim, we consider the paradigmatic Susceptible-Infectious-Recovered (SIR) model:
in this model, a Susceptible (S) node $i$ in contact with an Infectious one $j$ can become Infectious (I). If the edge between $i$ and $j$
has weight $W_{ij}$, the rate at which this event occurs is given by
$\beta W_{ij}/ T$ where $T$ is the total measurement time (see Methods and Stehl\'e et al. \cite{Stehle:2011a} for details). 
Each Infectious node becomes Recovered (R) at rate $\mu$ and cannot be infected anymore. The process ends 
when there are no Infectious nodes any more. We perform numerical
simulations of this model on the contact network, on the friendship network and on networks obtained by sampling the contact network
using several sampling methods (described in the next paragraph). 
To quantify the epidemic risk and compare outcomes of these simulations, we measure the distributions of epidemic sizes 
(i.e., the final fraction of recovered
nodes), the fraction of epidemics with size larger than $20\%$ and the average
size of these epidemics (the cut-off of $20\%$ is chosen arbitrarily to 
distinguish between small and large epidemics; changing the value of this threshold does not alter our results).

It is important to note that we consider here a static version of the contact network, while the data as hand provides
temporally resolved contacts. The rationale behind this choice is twofold. First, the friendship survey data 
does not contain temporal information. If using friendship network can be seen as a sampling of contact networks, it is thus 
necessarily a static sample. Second, when modeling the propagation of infectious diseases with realistic timescales of several days,
it has been shown in \cite{Stehle:2011a} that a static weighted contact network contains enough information 
to obtain a good estimate of the process outcome. Clearly, when dealing with faster processes, the temporal evolution of the network
becomes relevant; in that case, studies such as \cite{Genois:2015} have shown how to 
build realistic surrogate timelines of contacts on weighted
networks, using the robustness of the distributions of the durations of single contact events and of the intervals
between successive contacts measured in different contexts.

\subsection*{Sampling methods}

\begin{figure}[!ht]
\centering
\includegraphics[width=0.8\textwidth]{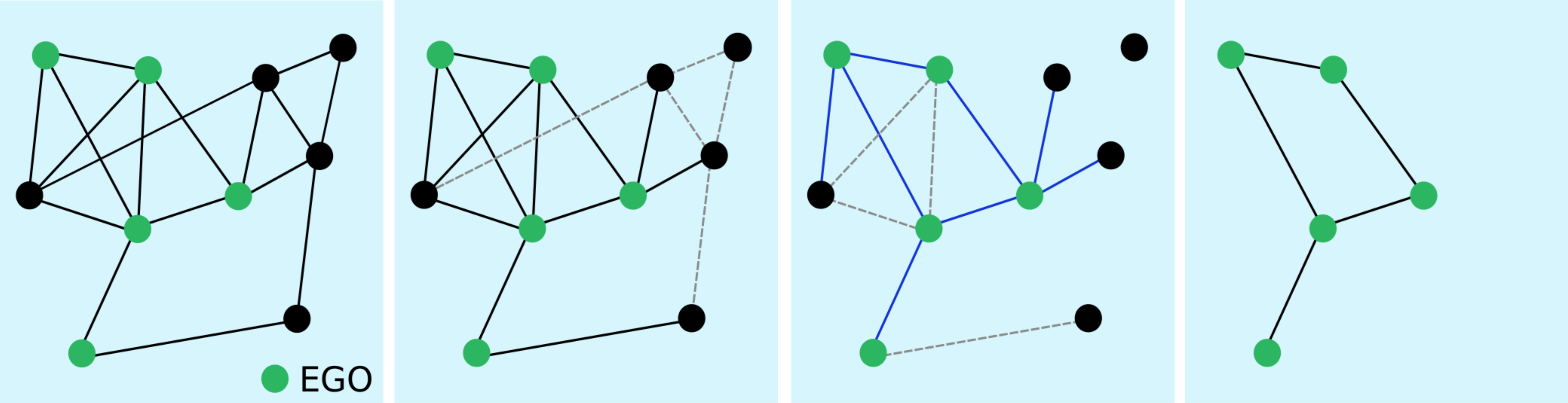}
\caption{\textbf{Sketch of the EGOref sampling process.} We first select a certain number of nodes as egos (in green), who represent
the respondents to the surveys. Links between non-respondents cannot be observed (dashed grey links in the second panel). Each ego then 
``chooses'' to report some of its links, with probability depending on their weights (links shown in blue in the third panel, while
the non-reported links are shown in dashed grey lines). 
We then finally keep only the egos and, among the chosen edges, only the ones joining egos (last panel).
 \label{fig:schema}}
\end{figure}

Many different sampling procedures of network data have been considered in previous works, and their impact on the 
network's statistical properties have been studied \cite{Granovetter:1976,Frank:1978,Achlioptas:2005,Lee:2006,Kossinets:2006,Onnela:2012,Blagus:2015}.
Sampling of the network used as substrate for transmission events 
is also known to affect the result of simulations of epidemic spread \cite{Ghani:1998a,Ghani:1998b,Genois:2015}.
In particular, population sampling has a strong impact, even in the case of uniform sampling \cite{Genois:2015}.
We therefore consider here the following sampling procedures on the contact network, tuned to obtain the same number of nodes as in the friendship network:
\begin{itemize}
\item We first consider as reference the Subgraph method (``SubFr''):
we consider the $134$ nodes of the Friendship network present in the contact network and take the subgraph induced by these nodes on 
the contact network. This would correspond to a population sampling of the contact network, with the sampled
population corresponding to the respondents of the friendship survey.

\item In the Random Node method (``RN''), we choose $N_F=135$ nodes uniformly at random from the contact network
and we take the subgraph induced by these nodes on the contact network. This corresponds to a population sampling
with uniformly random choice of the sampled nodes.

\item We also consider the Egocentric sampling method (``EGO''): 
we select a node at random and include this node and all its neighbors in the sample. 
We repeat this step until we reach the desired number of nodes, $N_F$. 
If this number is exceeded by including the chosen node and its neighbors, 
we randomly choose a set of its neighbors so that the right number of nodes is exactly reached. 
Then, we take the subgraph induced by this sample of nodes on the original network.
\end{itemize}
As these methods do not allow us to control the number of edges in the sampled network, we also consider
several additional sampling methods, in which we can tune this number and set it equal to $E_F$.
\begin{itemize}
\item In the Random Edge method (``RE''), we first choose edges at random from the contact network until we reach the desired number of nodes
$N_F$; as the number of chosen edges is still lower than $E_F$ we then
choose edges at random from the contact network with the condition that both their extremities are in the set of nodes obtained in the first step, 
until the desired number of edges is reached. 

\item In the Refined Random Node method (``RNref''), we add the following step to the RN method: after the subgraph is obtained, we remove edges at
random to get the desired number of edges in the final sampled network. 

\item  We propose a new Refined Egocentric method (``EGOref''), inspired
by the result of Ref \cite{Mastrandrea:2015} that the longest contacts 
corresponded to reported friendships, while many short contacts did not. Here, we select $N$ nodes called
\textit{egos} at random from the contact network. For each \textit{ego} $i$ we select some of its edges as follows: each edge
$i-j$ is selected with a probability equal to $\min\left( p*\frac{W_{ij}}{S_{i}},1 \right)$,
with $W_{ij}$ the weight of the edge between $i$ and $j$, $S_i = \sum_\ell W_{i\ell}$ the
strength of the \textit{ego} node $i$ and $p$ is the parameter of the
model.  We then keep only the \textit{egos} and the selected edges
linking them and we remove the other edges (between egos and non-egos) and nodes (non-egos). With this
method, we end up with the desired number of nodes by setting $N=N_F$, and a 
number of edges that depends on the parameter $p$. Figure \ref{fig:schema} summarizes this process. 

\item While the \textit{egos} are chosen uniformly at random among the nodes of the contact network
in the EGOref method, we also consider a heterogenous EGOref
method (``EGOref-het''), in which the distribution of \textit{egos} in the various high school classes corresponds
to the one of the friendship network  (\textit{egos} still being chosen at random within each class).
\end{itemize}

When using the sampled contact networks and the friendship network
to simulate SIR processes, we moreover assign to each sampled edge 
a weight 
taken at random 
from the empirical distribution of weights of the contact network, which is known to be a robust feature of human contact patterns 
\cite{Fournet:2014,Barrat:2015} (See also the Supplementary Information for more details about the assignment of weights to links).

\subsection*{Properties of sampled networks and outcomes of SIR simulations}

Table \ref{tab:prop1} shows the characteristics of the empirical contact and friendship networks
compared to the networks obtained by the simplest sampling techniques (SubFr, RN, EGO, RE).
The contact network has $327$ nodes, while the friendship and sampled networks have $135$ nodes. 
The density is twice higher in the contact network than in the friendship network; however, the 
subgraph induced by the nodes of the friendship network (SubFr) has in fact a slightly higher density than the contact network.
As the RN method samples uniformly the nodes of the contact network, the resulting density is on average equal to the density of the contact network.
On the other hand, the density of the EGO sampled networks is even higher. 
Finally, RE sampling yields by construction the same density as the friendship network.

The clustering coefficient displays interesting features: despite a much lower density, 
the friendship network has a higher clustering coefficient than the contact network.
Networks obtained through the SubFr, RN and EGO methods have as well rather large clustering coefficients, 
while the RE sampling yields much lower values.

\begin{table}[!ht]
\begin{tabular}{|l||l|l|l|l|l|l|l|l|}
\hline
& Number of nodes & Number of edges & Density & Average degree & Average clustering & Avg shortest path*\\
\hline
Contact network & 327 & 5818 & 0.11 & 35.6 & 0.503 & 2.15 \\
Friendship network & 135 & 413 & 0.05 & 6.1 & 0.532 & 4.06 \\
\hline
SubFr & 134 & 1235 & 0.14 & 18.4 & 0.546 & 2.22 \\
RN & 135 & 987 & 0.11 & 14.6 & 0.498 & 2.37 \\
EGO & 135 & 1679 & 0.19 & 24.9 & 0.573 & 2.04 \\
\hline
RE & 135 & 413 & 0.05 & 6.1 & 0.157 & 3.19 \\
\hline
\end{tabular}
\caption{Features of the empirical networks and of the networks obtained with the simplest sampling methods:
SubFr, RN, EGO preserving the number of nodes
and RE preserving both the number of nodes and edges 
of the friendship network. *The average shortest path length is computed on the largest connected component of the network.
\label{tab:prop1}}
\end{table}

\begin{figure}[!ht]
\centering
\includegraphics[width=\textwidth]{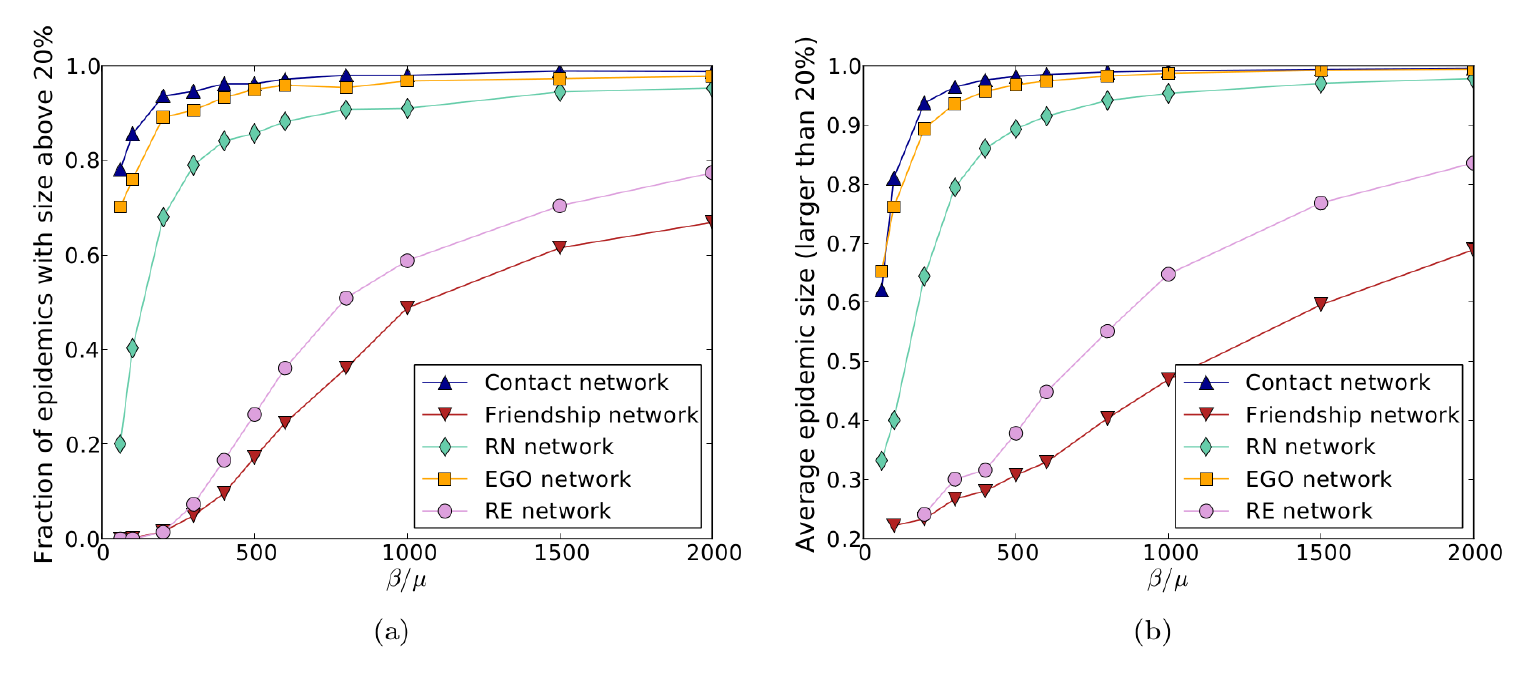}
\caption{\textbf{Outcome of SIR spreading simulations performed on empirical and sampled networks.} 
(\textbf{a}) Fraction of epidemics with size above 20\% (at least 20\% of recovered individuals at the end of the SIR process) as a function of 
$\beta/\mu$. 
(\textbf{b}) Average size of epidemic with size above 20\% as a function of the parameter of spreading $\beta/\mu$.
\label{fig:epi1}}
\end{figure}

Figure \ref{fig:epi1} shows the outcome of the spreading simulations 
performed on the two empirical networks and on sampled networks obtained by the RN, RE and EGO sampling methods: it displays
the fraction of epidemics with size above $20\%$ and the average size of epidemics among the ones with size above $20\%$, as a function 
of the spreading parameter $\beta/\mu$. As expected, simulations using the friendship network
give a very strong underestimation of the epidemic risk with respect to the ones using the contact network. 
Moreover, simulations performed on the sampled networks RN and EGO yield a much
larger estimation of the epidemic risk than when using the friendship network,
and this estimation increases as the density of the network obtained by sampling increases (as is expected since the 
availability of transmission paths increases). Finally,
simulations using the RE sampled networks also yield larger estimated epidemic sizes than when 
using the friendship network, although these networks
have the same density. This is in agreement with the results of Smieszek et al.\cite{Smieszek:2009} stating 
that high clustering values tend to hinder propagation processes, at fixed density: 
the clustering coefficient of the RE networks is indeed much smaller than the one of the friendship network.

We show in the Supplementary Information the results of simulations performed on the SubFr network as well as on a randomized version
of the friendship network using the rewiring algorithm of Maslov et al.\cite{Maslov:2004} (MSZ). The SubFr network corresponds to a
population sampling of the contact network, and leads to a limited underestimation of the 
epidemic risk with respect to the whole contact network. The MSZ network has the same number of nodes and edges than the frienship network, hence 
the same density, but a much smaller clustering, due to the randomization, and gives thus a higher epidemic risk.


\begin{figure}[!ht]
\centering
\includegraphics[width=0.4\textwidth]{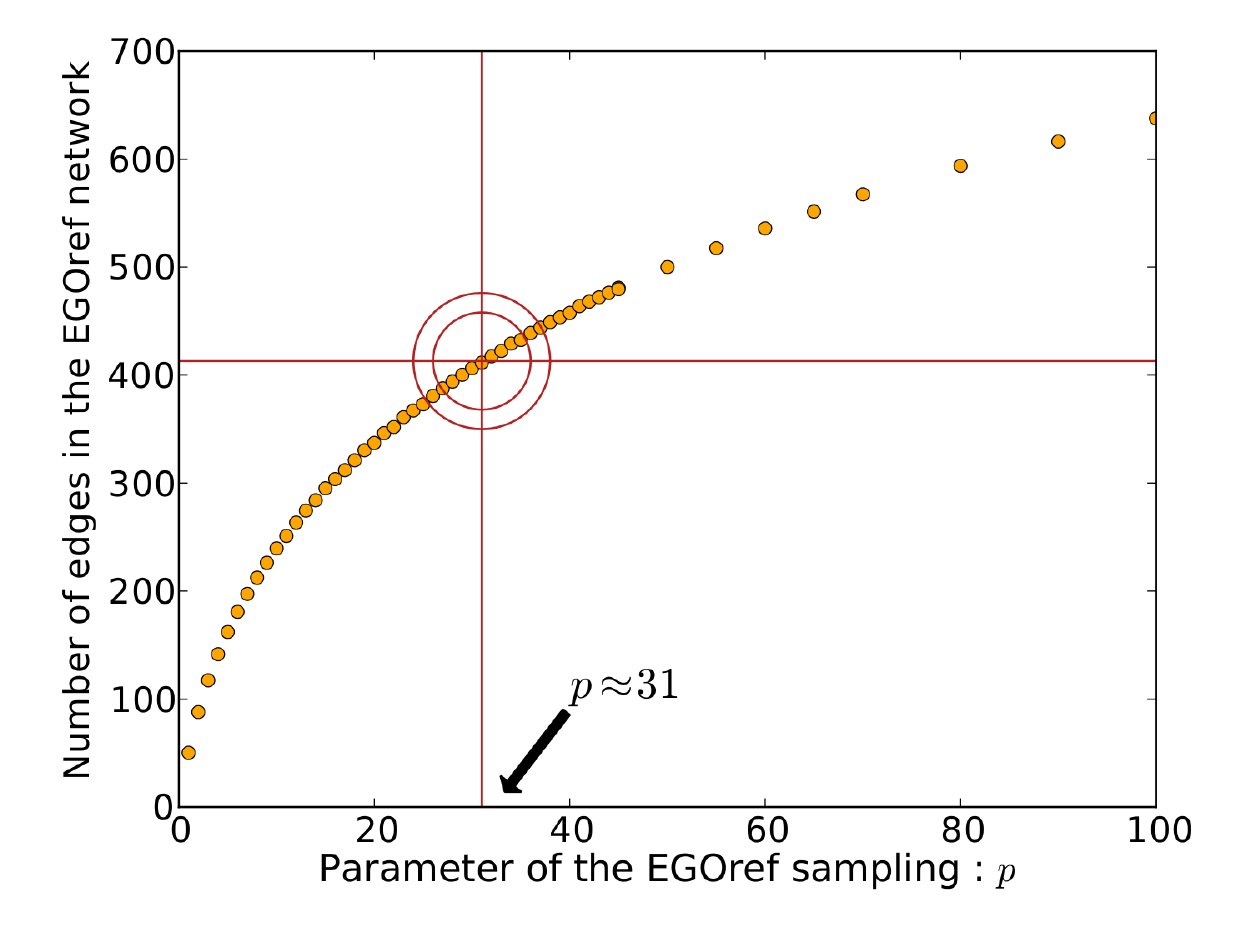}
\caption{\textbf{Average number of edges in the sampled network obtained by the EGOref method as a function 
of the parameter $p$ with $N_F=135$.} 
The horizontal red line represents the number of edges $E_F$ in the friendship network.\label{pnbre}}
\end{figure}

We now turn to the case of the EGOref sampling method. In the resulting sampled networks,
the number of edges depends on the parameter $p$, at fixed number of sampled nodes $N_F$. Figure \ref{pnbre}
shows the average number of edges in the sampled network as a function of $p$. For $p > \max_{i,j} ( s_i/W_{ij}) $, this number
reaches the average number of edges in the subgraph induced by $N_F$ nodes chosen at random on the contact network, which is equal
to the number of edges obtained through the RN sampling process. As our goal is 
to obtain a sampled subgraph of the contact network that is similar to the friendship network, we
tune $p$ in order to obtain sampled networks with an averaged number of edges close to $E_F=413$, the number of edges in the friendship network.
This value is obtained for $p \approx 31.3$ (Figure \ref{pnbre}). At this value, the obtained network results always connected.

\begin{table}[!ht]
\begin{tabular}{|l||l|l|l|l|l|l|}
\hline
& Number of nodes & Number of edges & Density & Average degree & Average clustering & Avg shortest path* \\
\hline
Friendship network & 135 & 413 & 0.05 & 6.1 & 0.532 & 4.06 \\
RE & 135 & 413 & 0.05 & 6.1 & 0.157 & 3.19 \\
RNref & 135 & 413 & 0.05 & 6.1 & 0.197 & 3.21 \\
EGOref & 135 & 413 & 0.05 & 6.1 & 0.355 & 3.90 \\
\hline
\end{tabular}
\caption{Features of the friendship network and of the sampled networks obtained by 
RE, RNref and EGOref method, all of them preserving the number of nodes and edges of the friendship network. 
*The average shortest path length is computed on the largest connected component of the network.
\label{tab2}}
\end{table}

Table \ref{tab2} compares the properties of the contact networks sampled by the RE, RNref and EGOref methods, in which we tune the number of edges,
with the friendship network. Both average shortest path and clustering coefficient are closest to the friendship network case for the EGOref method,
but the clustering remains in all cases smaller in the sampled networks than in the friendship network. 
Further refinements of the EGOref method might yield a clustering closer to the
one of the friendship network, at the cost however of an increase in the method's complexity and number of parameters.

\begin{figure}[!ht]
\centering
\includegraphics[width=.7\textwidth]{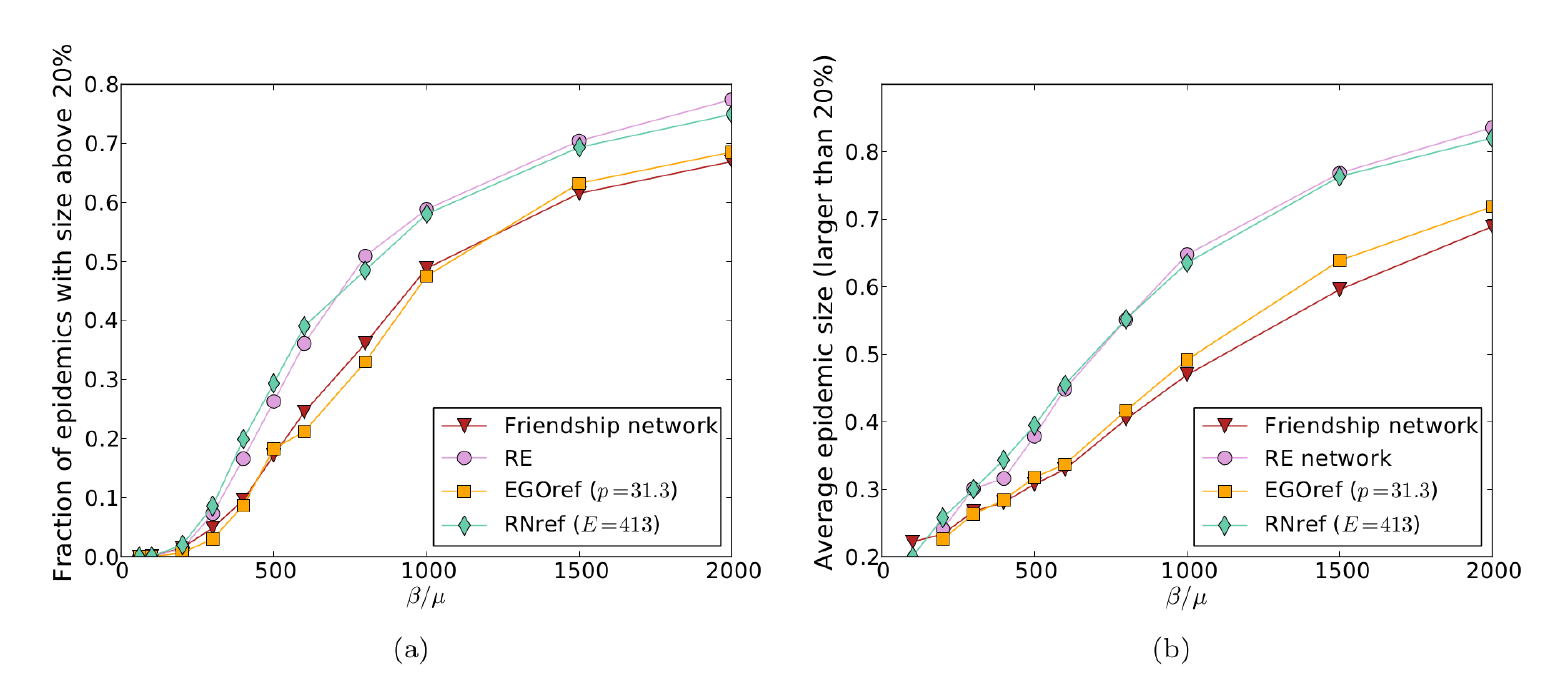}
\caption{\textbf{Outcome of SIR spreading simulations performed on friendship and sampled networks.} 
(\textbf{a}) Fraction of epidemics with size above 20\% (at least 20\% of recovered individuals at the end of the SIR process) as a function of the parameter of spreading $\beta/\mu$. 
(\textbf{b}) Average size of epidemic with size above 20\% as a function of the spreading parameter $\beta/\mu$. 
\label{egorefthebest}}
\end{figure}

\begin{figure}[!ht]
\centering
\includegraphics[width=.7\textwidth]{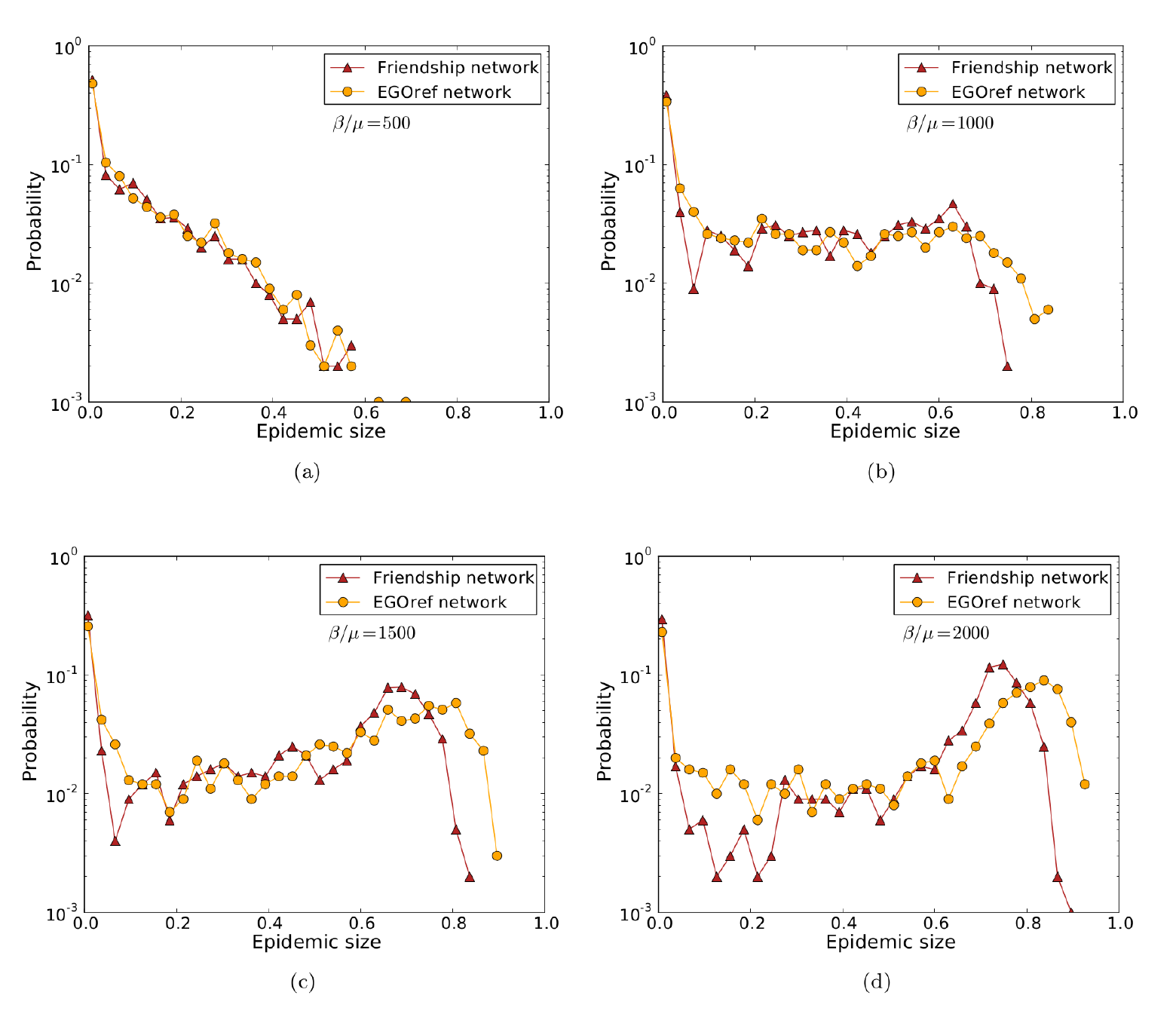}
\caption{\textbf{Distributions of epidemic sizes of SIR spreading simulations.}
We compare the outcomes of SIR spreading simulations performed on friendship and EGOref networks for different  
values of $\beta/\mu$.
\label{egovsfr}}
\end{figure}

Figure \ref{egorefthebest} shows the outcomes of epidemic spreading simulations
performed on the friendship network and on the contact networks sampled using the EGOref, RNref and RE methods. 
A very good agreement with the epidemic risk estimated from the friendship network is obtained for the EGOref sampling, while
the RE and RNref sampled networks yield large epidemic sizes with higher probability and larger average epidemic sizes,
even if they have the same density. As mentioned above, this could be expected given their smaller clustering. 

Figure \ref{egovsfr} displays the whole distributions of epidemic sizes for simulations performed on the friendship and EGOref networks, for
4 values of the spreading parameter $\beta/\mu$. A good agreement in the shape of the distributions is observed, although
the maximal size of epidemics is systematically higher in the EGOref sampled contact networks 
than in the friendship network, especially at large ${\beta}/{\mu}$.

\begin{figure}[!ht]
\centering
\includegraphics[width=.6\textwidth]{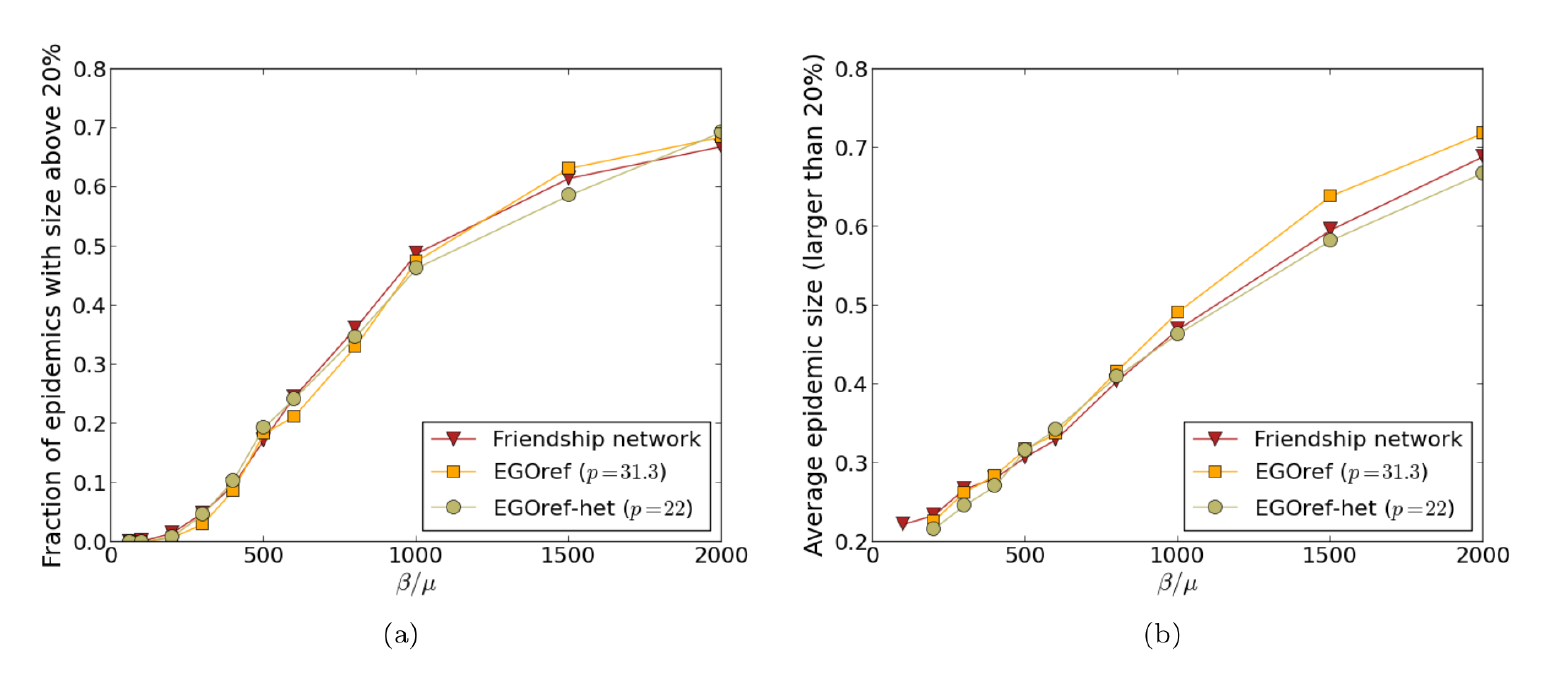}
\caption{\textbf{Outcome of SIR spreading simulations performed on friendship and both EGOref sampled networks.} (\textbf{a}) Fraction of epidemics with size above 20\% (at least 20\% of recovered individuals at the end of the SIR process) as a function of the parameter of spreading $\beta/\mu$. (\textbf{b}) Average size of epidemic with size above 20\% as a function of the spreading parameter $\beta/\mu$.\label{ego22}}
\end{figure}

\begin{figure}[!ht]
\centering
\includegraphics[width=.6\textwidth]{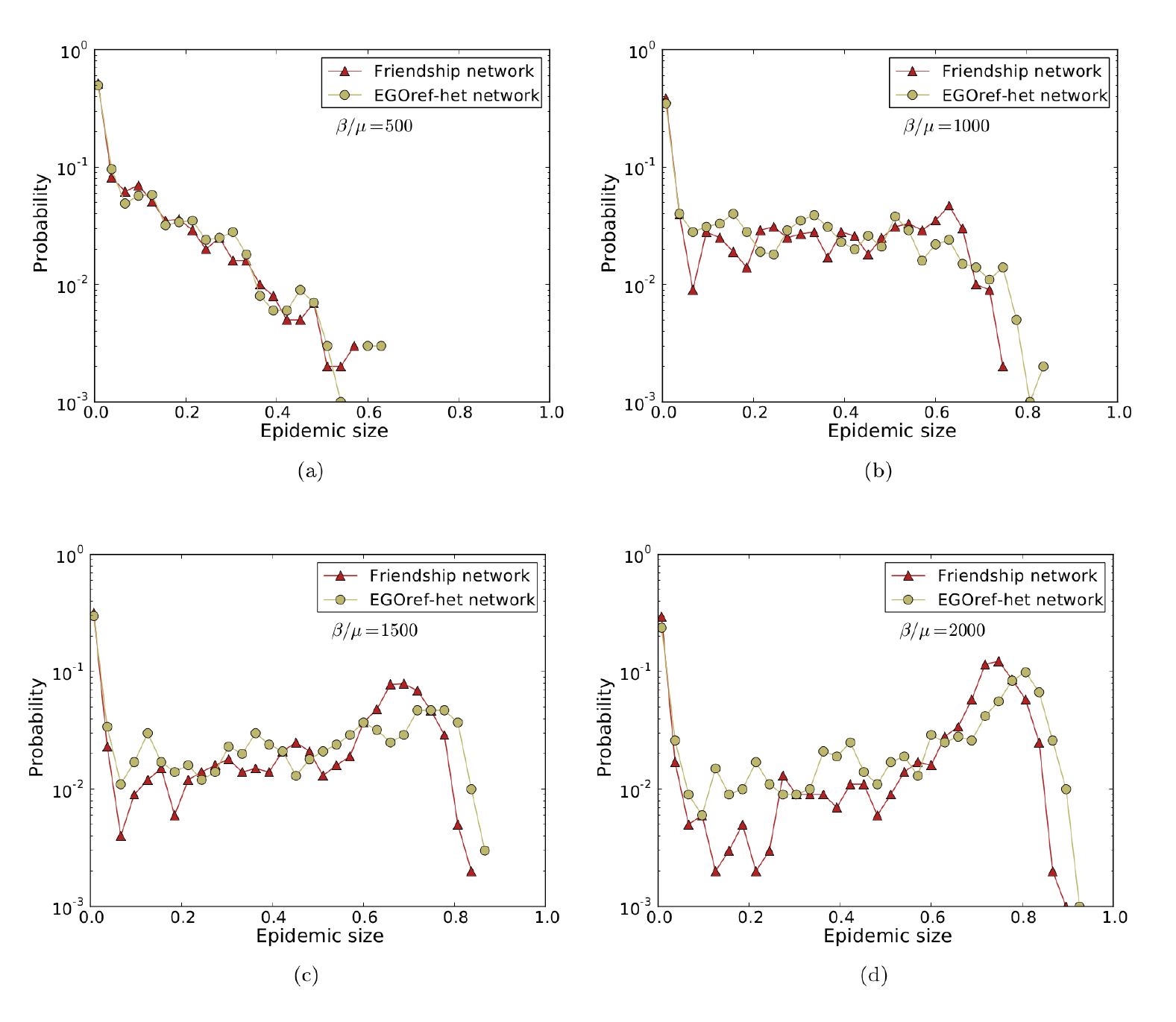}
\caption{\textbf{Comparison of the distributions of epidemic sizes of SIR spreading simulations performed on friendship and 
heterogeneous EGOref sampled networks for different values of $\beta/\mu$.}
\label{egovsfr2}}
\end{figure}

We finally consider the EGOref-het sampling method. In that case, the number of sampled nodes is equal to the number of nodes of the friendship network
in each class. We fix the number of edges to $413$ by the same method as in the EGOref case; this is obtained
for $p \approx 22$. The resulting average shortest path length is $4.03$ and the average clustering is $0.334$, still smaller than in the friendship network and
very close to the clustering in the EGOref sampled networks. 
We show in Figures \ref{ego22} and \ref{egovsfr2} that the outcomes of the spreading simulations are similar to the EGOref case, with a slightly
better agreement with the results of simulations using the friendship network, in particular for large epidemics.

\subsection*{Sampling model exploration}

In this section, we investigate how simulations of spreading processes performed on networks
obtained from the contact network using the EGOref sampling method depend on the method's parameters $p$ and $N$. 
We consider $135 \le N \le 327$ and $15 \le p \le 500$ (as for $p>500$ the number of links is almost equal to its maximum possible value). 
We first observe (not shown) that, at fixed $N$, the density of the sampled network is fully determined by the value of $p$. Changing $p$ is thus
equivalent to tuning the resulting network's density. For $p=500$ and $N=327$, we almost recover the whole contact network.

\begin{figure}[!ht]
\centering
\includegraphics[width=.7\textwidth]{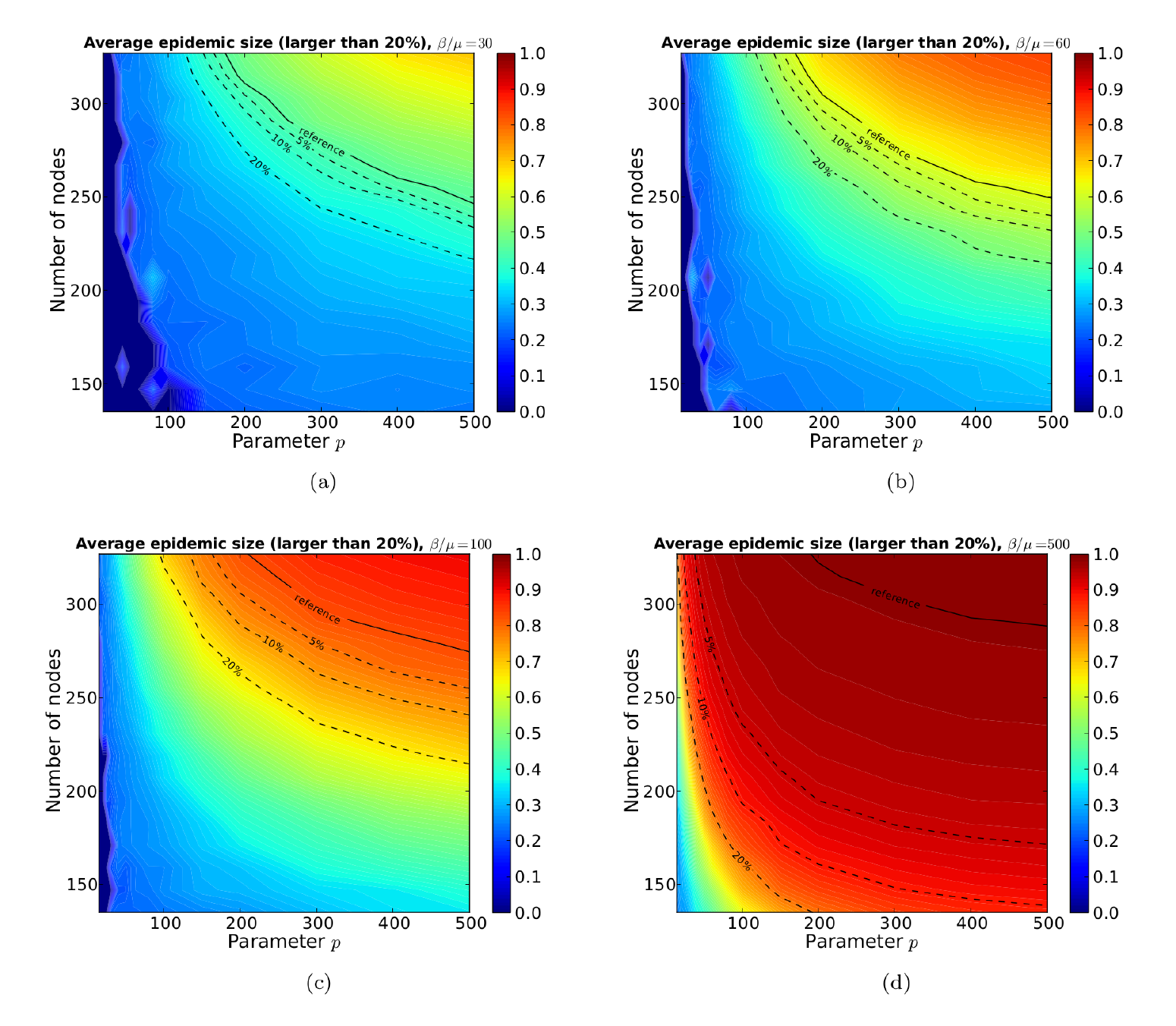}
\caption{\textbf{Color maps of the average epidemic size for epidemics with size above 20\% for several 
values of $\beta/\mu$.}
When no epidemics has size above 20\%, the value is zero. The three dashed lines represent the value of 
average epidemic size at 5\%, 10\%, 20\% of the reference value (solid line), 
which corresponds to the average epidemic size of the SIR spreading simulations performed on the contact network.\label{colorplot}}
\end{figure}

In Figure \ref{colorplot}, we show the average epidemic size obtained on the sampled networks as a function of $p$ and $N$ for different values 
of the spreading parameter $\beta/\mu$: this size increases with both $p$ and $N$. Increasing $p$ (which determines the link reporting probability)
at fixed $N$ (the number of survey participants) or the contrary is not enough to obtain a correct estimation of the epidemic risk: both
have to be increased in order to obtain the same value as when using the whole contact network, shown by
the continuous line. 
The dashed lines show the values of $p$ and $N$ necessary to obtain an estimation of the epidemic size within
5\%, 10\% or 20\% of this reference value.

We also note that the average epidemic size obtained for the largest values of $p$ and $N$ is actually larger than the reference, although the 
corresponding EGOref network is structurally almost the same as the contact network. This discrepancy stems from the fact that the 
edge weights are placed differently in both cases: weights are indeed assigned at random on the edges of the network obtained by the
EGOref sampling procedure. We show in the Supplementary Information that a weight reshuffling on the contact network leads indeed to an increase in the
size of the simulated epidemics. 

We moreover show in the Supplementary Information the impact of the EGOref sampling on the outcome of SIR simulations 
for two other data sets corresponding to contacts in a conference and in offices, comparing it with the result of a simple population sampling.
The underreporting of contacts as modeled by $p$ is shown to have a very strong impact.

\section*{Discussion}

In this paper, we investigated if data coming from friendship surveys can be considered as similar 
to a sampling procedure on the contact network in a population, in the context of the estimation of the outcome 
of spreading processes. 
We focus here on infectious diseases for which contact networks are considered as a relevant
proxy of transmission possibilities \cite{Salathe:2010}.  
The rationale leading to this question comes from the quantitative comparison of friendship and contact networks collected within the same population,
discussed in Mastrandrea et al.\cite{Mastrandrea:2015} in the context of a high school:  
friendship and contact networks are indeed different, and many short contacts occur between individuals who are not friends; however, 
the longest contacts, which play an important role in potential propagation events, effectively correspond
to friendship links, and the overall structure of the networks, in terms of interactions between different classes, are similar.
Objectively measured contact networks are not always available, and 
friendship networks could be easier to obtain than contact networks in some situations. Moreover, 
surveys asking individuals about their friends might suffer less from memory biases than contact diaries. Understanding
if and how friendship surveys could be used in models of spreading processes would thus be interesting. 
Most importantly, framing the relation between friendship networks and actual contact networks as a sampling procedure
might also help design and evaluate procedures to compensate for the resulting biases in the estimation of epidemic risks.

To make progresses in this direction, we have considered a publicly available data set combining, for the same set of individuals, a 
contact network measured by wearable sensors and a friendship network obtained by a survey. We have considered
several ways of sampling the contact network and investigated their similarity with the friendship network with respect to simulations of epidemic spread.
We have first considered a uniform random sampling of nodes, as the friendship network has much less nodes than the contact network.
The RN method simply samples nodes at random and considers the resulting induced subgraph, while the EGO method is closer to
a survey procedure, as it includes nodes chosen at random and all their contacts.
In both cases, the obtained networks have much higher densities than the friendship network. As a consequence, the outcomes of epidemic spreading
performed on these networks yield a much higher estimation of the epidemic risk than the friendship network. 
We have therefore considered sampling methods in which the density could be tuned, at fixed number of nodes. 
In the RE and RNref methods, we choose edges at random, which leads to a rather small clustering coefficient in the sampled networks and thus
the resulting estimation of the epidemic risk is still higher than for the friendship network. These methods
seem indeed too naive to recover the small scale structures and correlations of the friendship network.

Finally, we designed the EGOref sampling method, in a way to mimic a survey procedure in which nodes (egos) report on their friendships, 
under the assumptions that all reported friendships correspond to contacts, and that the probability that a contact 
corresponds to a friendship is larger for longer contacts. Note that this method is not aimed at reproducing exactly the friendship
network (in which some links in fact do not correspond to contacts), but its goal is to produce a sampled network whose properties
are close enough to the friendship network to lead to similar outcomes when used in simulations of spreading processes. 
The sampling method depends, at fixed number of egos, on a single parameter $p$, which determines the density of the resulting sampled network. 
In particular, for a uniform sampling of nodes, we can choose $p$ to have the same density as the friendship network. 
At fixed density, this method allows us to recover a higher coefficient of clustering than the other sampling methods. 
Most importantly, simulations of spreading processes performed on the resulting sampled network yield an estimation of the epidemic risk
in very good agreement with simulations using the friendship network, for a wide range of spreading parameters.

Some limitations of our approach are worth discussing. First, the EGOref method considers a uniform sampling of nodes. 
Choosing the \textit{egos} with the same class distributions than in the friendship network leads to a slightly better agreement, as
shown through the EGOref-het method. The clustering coefficient of the sampled networks is however still lower than in the empirical friendship network.
More involved sampling procedures allowing to control the clustering might be sought, but would result in more complex and less intuitive sampling rules.
The way we choose to assign the weights to edges in the sampled networks can also be discussed. 
On the one hand, the distribution of weights is known to be very robust in human contact networks, even in very different contexts 
such as schools or hospitals \cite{Stehle:2011a,Vanhems:2013}, so that it seems natural to  use the empirical distribution of weights,
which can be taken from publicly available datasets, to assign weights to links obtained through surveys.
On the other hand, a random assignment of weights destroys edge-weights correlations that can have an impact
on the outcome of spreading simulations. 
We have considered such a random assignment as our goal is to compare the friendship network to a sampled contact network: when only friendship data is 
available, a natural way to perform simulations of epidemic spreading processes is indeed to assign weights at random to the friendship links. 
More accurate ways to assign weights in order to mimic the correlations linking the strengths and degrees of nodes in the contact network
would be of great interest, but might depend on the context. 
We have also considered static networks, while real contact networks evolve over time. As mentioned above, this is based on a twofold rationale:
as friendship data to not include temporal information, the comparison should be done with a sampling of a static weighted contact network; Moreover,
for slow propagation timescales corresponding e.g. to flu-like illnesses, 
the precise dynamic of the contacts does not represent a crucial information \cite{Stehle:2011a}. In order to extend
the comparison to faster processes, one should create surrogate timescales on the links similarly to Ref.\cite{Genois:2015}.
Finally, our study is based on only one specific population. This is due to the current lack of datasets in which both contact and friendship networks
are available. Further investigations in different contexts would be of great interest.

The sampling method we have proposed here depends on two parameters: the number of respondents $N$ and the parameter $p$ 
that allows to tune the amount of contacts reported. We have started here to study its impact on 
 various data sets of contact networks in different contexts, and future work will focus on designing and evaluating 
methods to estimate the epidemic risk from incompletely and non-uniformly sampled data,
such as the one described by G\'enois et al. \cite{Genois:2015} for uniform population sampling, and to 
understand their efficiency and limits when the sampling parameters are varied.

\section*{Methods}
\subsection*{Data}
The data we consider was collected by the SocioPatterns collaboration in a French high school in Marseille during the week December 2nd - 6th 2013,
and subsequently analysed and made publicly available \cite{Mastrandrea:2015}. We use here two different data sets
describing different types of relationships between students. 
The first corresponds to close face-to-face interactions between individuals equipped with wearable sensors. These 
contacts were measured for 327 students structured in 9 classes (called "classes pr\'eparatoires") corresponding to different
fields of study: 3 classes of Biology with respectively 37, 33 and 40 students, 
3 classes of Mathematics-Physics with respectively 33, 29 and 38 students, 
3 classes of Physics-Chemistry with respectively 44, 39 and 34 students. 
The participation rate reached 86.3\% (there were overall 379 students in the 9 classes).
The second type of data describes friendships relations between students. 
Students were asked to give the names of their friends within the high school. 
Over the 327 participants, only 135 students (41.3\%) answered the survey. 
Students who were cited as friend of a respondent but did not answer the survey were removed from the data. 
The 135 participants to the survey are distributed in the 9 classes as following: 10, 20 and 28 participants in the 3 classes of Biology; 
21, 3 and 7 participants in the 3 classes of Mathematics-Physics; 
21, 10 and 15 participants in the 3 classes of Physics-Chemistry. A link is drawn in the friendship network between two students if at least one
reported the other as friend.

\subsection*{SIR simulations}
To perform epidemic spreading simulations on the different networks, we need to take the link weights into account. 
The weight $W_{ij}$ of the edge between nodes $i$ and $j$ of the contact network gives the total duration of the contacts
that occurred between $i$ and $j$ during the total duration of the measure, $T$. We thus consider normalised weights 
$w_{ij}=W_{ij}/T$: for a Susceptible node $i$ in contact with an Infectious node $j$, the
probability of becoming Infectious is $\beta\cdot w_{ij}\cdot dt$ per time step $dt$.
For an Infectious node, the probability of becoming Recovered during the time step $dt$ is $\mu\cdot dt$. 
Each process starts with all nodes in the susceptible state except one infectious chosen at random (the seed).
As we consider static networks, the parameters $\beta$ and $\mu$ enter only through their ratio $\beta/\mu$ for the determination
of the outbreak size.  For each value of $\beta/\mu$, results are averaged over 1000 simulations with randomly chosen seed. 
For the simulations on the friendship network, the weights are assigned at random to the links for each run.
For the simulations on the sampled networks, a different sampling and weights assignment are performed for each run.

\section*{Acknowledgements}
A.B. is partially supported by the A*MIDEX project (ANR-11-IDEX-0001-02) funded by the ''Investissements d'Avenir''
French Government program, managed by the French National Research Agency (ANR), 
by the French ANR project HarMS-flu (ANR-12-MONU-0018), 
and by the EU FET project Multiplex 317532.

\section*{Author contributions statement}
A.B. and J.F. conceived and designed the study. J.F. performed the numerical simulations and the statistical analysis.
A:B. and J.F. wrote the manuscript.

\section*{Competing financial interests}
 The authors declare no competing financial interests.

\newpage

\setcounter{figure}{0}
\renewcommand{\thefigure}{S\arabic{figure}}

\setcounter{table}{0}
\renewcommand{\thetable}{S\arabic{table}}

\renewcommand{\thesection}{S\arabic{section}}

\begin{center}
{\Large
\textbf{Epidemic risk from friendship network data: an equivalence with a non-uniform sampling of contact networks.
Supplementary Information}
}\\
\vskip .3cm
\end{center}

\section{Cases of the SubFr and the randomized friendship network}

\begin{figure}[!ht]
\centering
\hspace{-10mm}
\subfigure[]
{\includegraphics[width=0.5\textwidth]{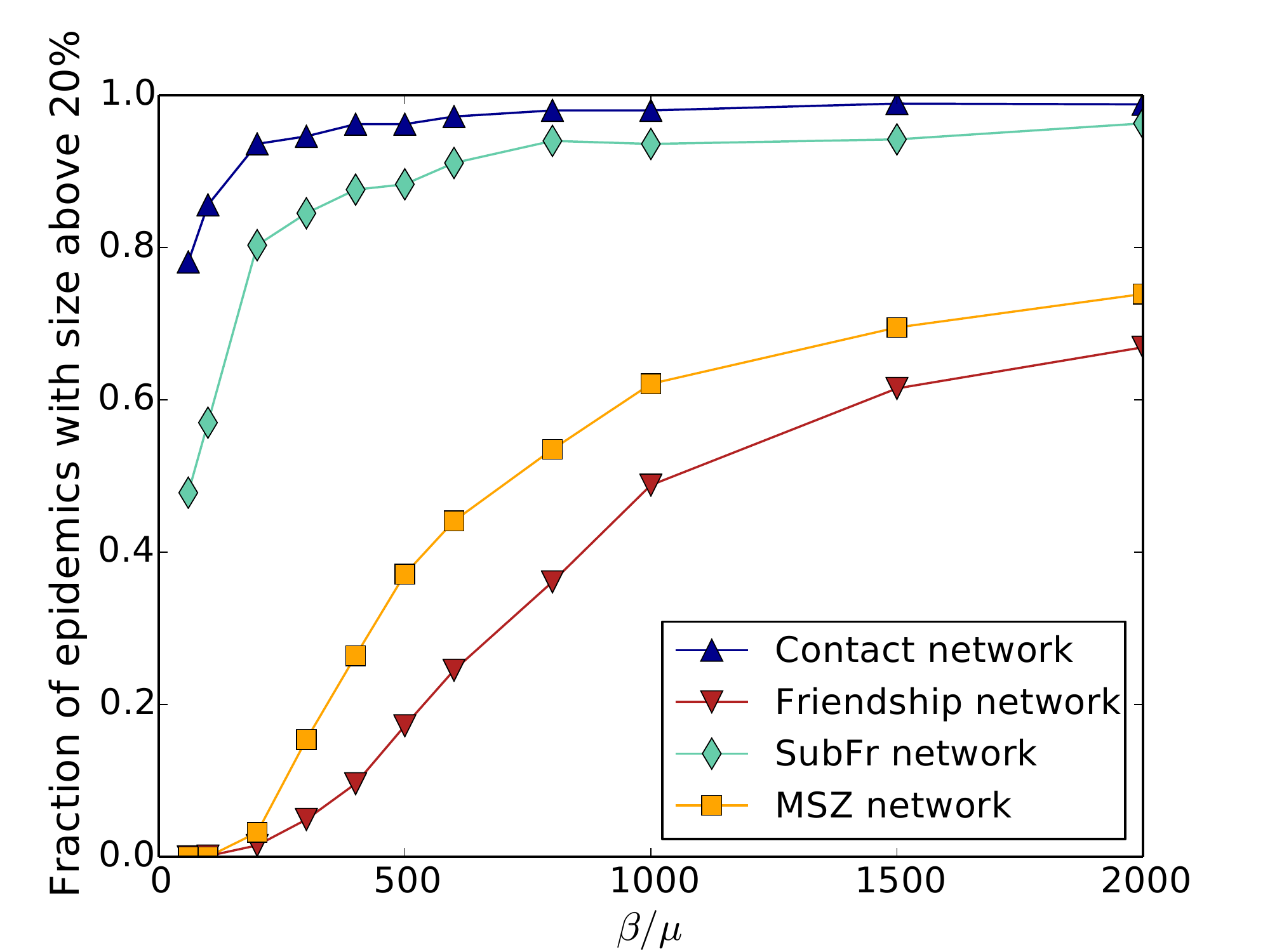}}
\subfigure[]
{\includegraphics[width=0.5\textwidth]{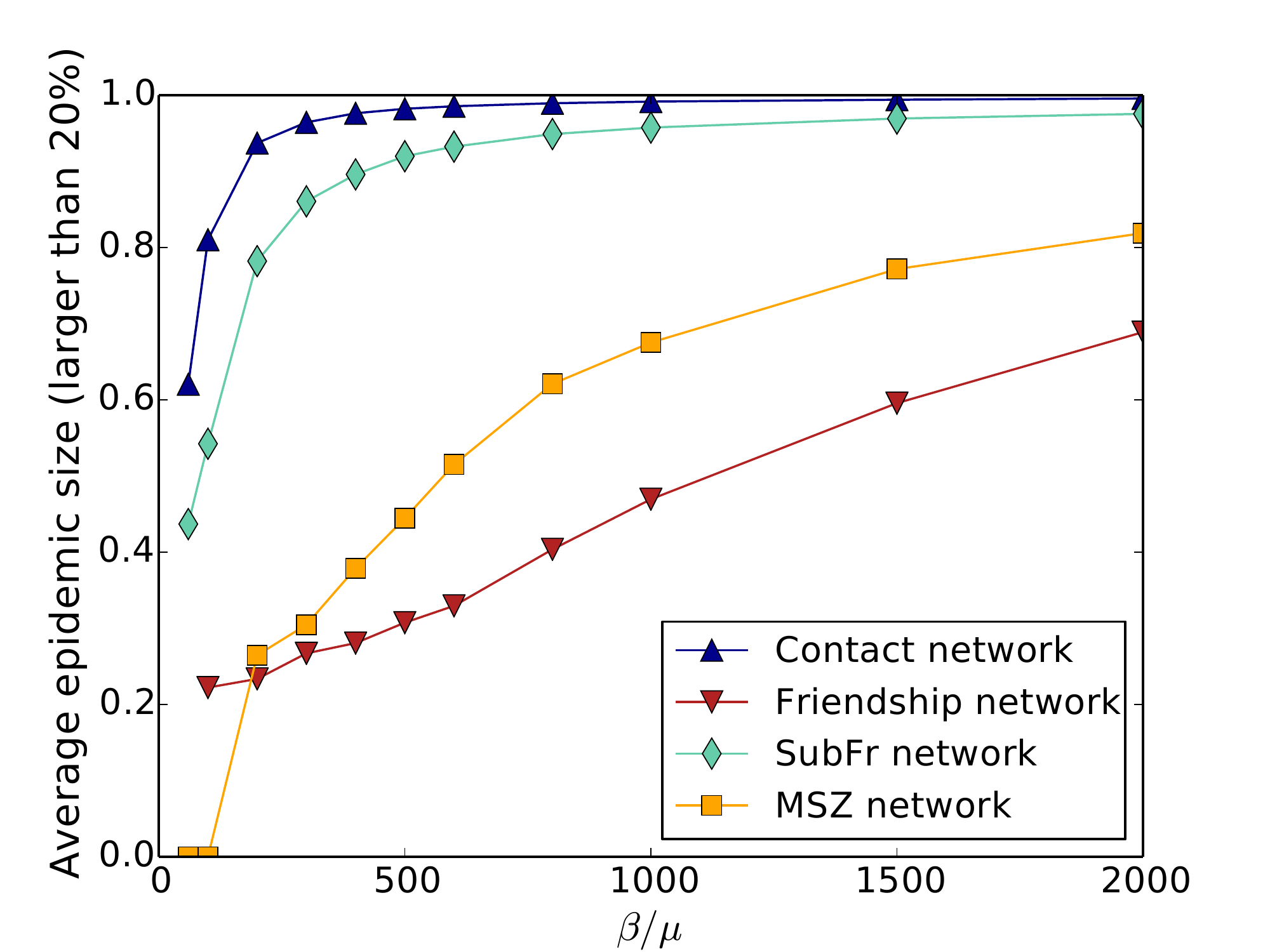}}
\caption{\textbf{Outcome of SIR spreading simulations performed on empirical, sampled and reshuffled networks.} 
We compare here the case of the SubFr sampling and of the randomized friendship network with the empirical contact and friendship networks.
({a}) Fraction of epidemics with size above 20\% (at least 20\% of recovered individuals at the end of the SIR process) as a function of 
$\beta/\mu$. ({b}) Average size of epidemics with size above 20\% as a function of $\beta/\mu$. \label{fig:s0}}
\end{figure}

We show in Fig. \ref{fig:s0} the outcome of SIR simulations performed on 
\begin{itemize}
\item the SubFr network, obtained as 
the subgraph induced by the nodes who participated to the friendship survey on
the contact network. This would correspond to a simple (non-uniform) population sampling of the contact network.
Due to this population sampling, the nodes' degrees are underestimated and the outcome of spreading processes leads to smaller
epidemic risk than when using the whole contact network. The difference is however rather contained with respect to
the difference between contact and friendship networks.

\item a randomized version of the friendship network using the algorithm of Maslov et al.\cite{Maslov:2004_2}. In this case the number
of nodes and links is exactly the same as in the friendship network, but structures and correlations are destroyed by the reshuffling. In particular,
the clustering coefficient is much smaller, favoring the propagation. The obtained epidemic risk is higher than for the friendship network, but much smaller than for the
contact network and the SubFr network.
\end{itemize}

\section{Assigning weights to links}

We investigate here the impact of several possible ways to assign weights to the networks used
for the simulation of spreading processes, namely the empirical contact network measured by the
wearable sensors, the network of reported friendships, and the networks obtained
through various sampling procedures from the contact network.
Indeed:
\begin{itemize}
\item the measured contact network is weighted, as the sensors give access to the duration of contacts.
We can therefore consider the weighted network with its true weights (``Original weights'') or, to assess
the impact of correlations between weights and structure, reshuffle them at random among the network's
links  (``Reshuffled weights'').

\item any sampling procedure produces a subgraph of the contact network. As a 
consequence, the weights of the edges can be either
taken directly from the contact network (``Original weights''), under the hypothesis
that the sampling procedure keeps information about the weights. On the other hand, the opposite
hypothesis, that sampling only informs about the existence of a link, and not on its importance, leads 
us to another weight assignment procedure, namely a random 
weight assignment from the distribution of weights of the contact network (``Random weights''). This is the procedure 
considered in the main text, as it is the most parsimonious and realistic in terms of availability of information.

\item the friendship network is not weighted. In order to use it in the simulations of SIR process, we can 
assign weights to links in different ways:
\begin{itemize}
\item we can choose the weights randomly from the distribution of weights
in the contact network (``Random weights'') or, 

\item for each edge of the friendship network present in
the contact network (86\% of the links in the friendship network find
a corresponding link in the contact network), we can use the corresponding
weight and, for the remaining 14\%, we can take the weights at random from the distribution 
of weights obtained from the first step (we call this assignment procedure ``Original weights''), or,

\item we assign the weights as in the previous method and then reshuffle randomly the weights
among the links of the friendship network (``Reshuffled weights''). 
\end{itemize}
\end{itemize}

\subsection{Contact network and sampling procedures independent from weights}

We note that methods of sampling in which edges are sampled independently from their weight (RE, RN, EGO) 
preserve the distribution of weights of the contact network. 
Moreover, reshuffling the weights in the contact network does not either change the distribution of weights.

\begin{figure}[!ht]
\centering
\hspace{-10mm}
\subfigure[]
{\includegraphics[width=0.35\textwidth]{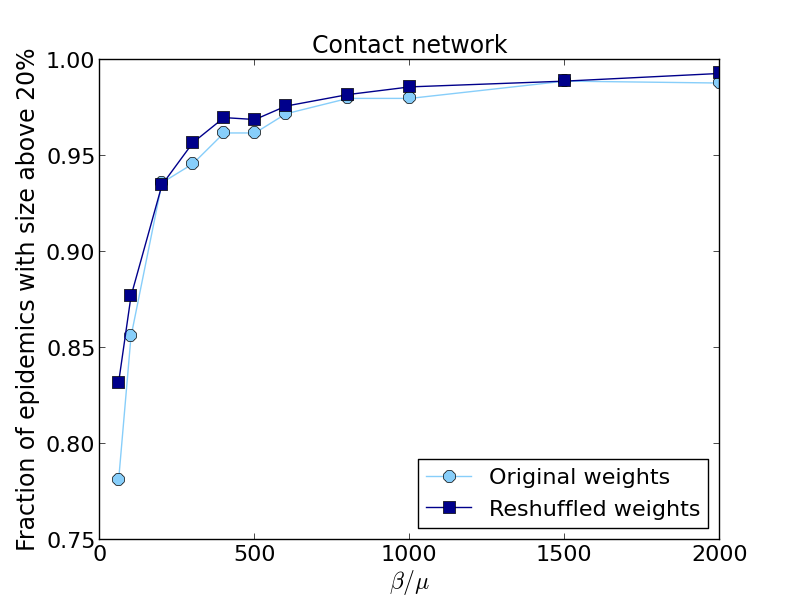}}
\subfigure[]
{\includegraphics[width=0.35\textwidth]{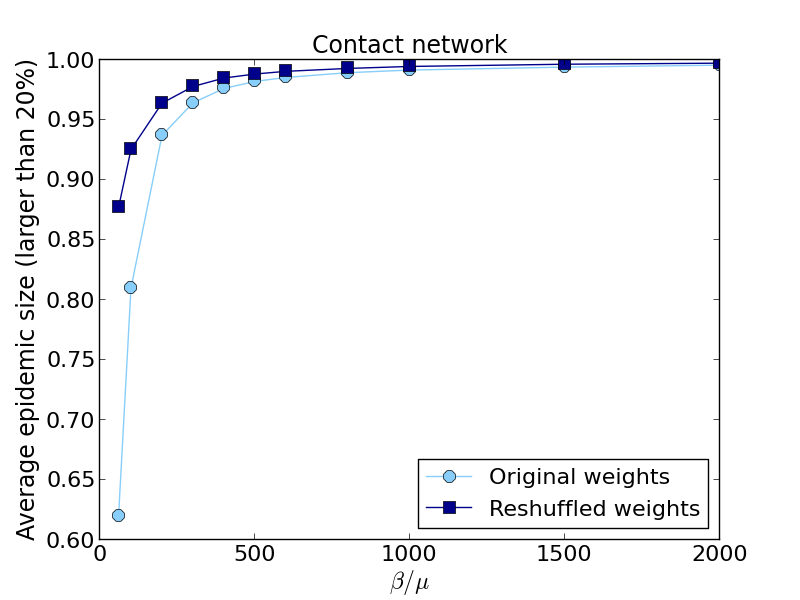}}
\caption{\textbf{Outcome of SIR spreading simulations performed on contact network with ``Original weights'' and ``Reshuffled weights''.} 
({a}) Fraction of epidemics with size above 20\% (at least 20\% of recovered individuals at the end of the SIR process) as a function of 
$\beta/\mu$. ({b}) Average size of epidemics with size above 20\% as a function of $\beta/\mu$. \label{fig:s1}}
\end{figure}

\begin{figure}[!ht]
\centering
\hspace{-10mm}
\subfigure[]
{\includegraphics[width=0.35\textwidth]{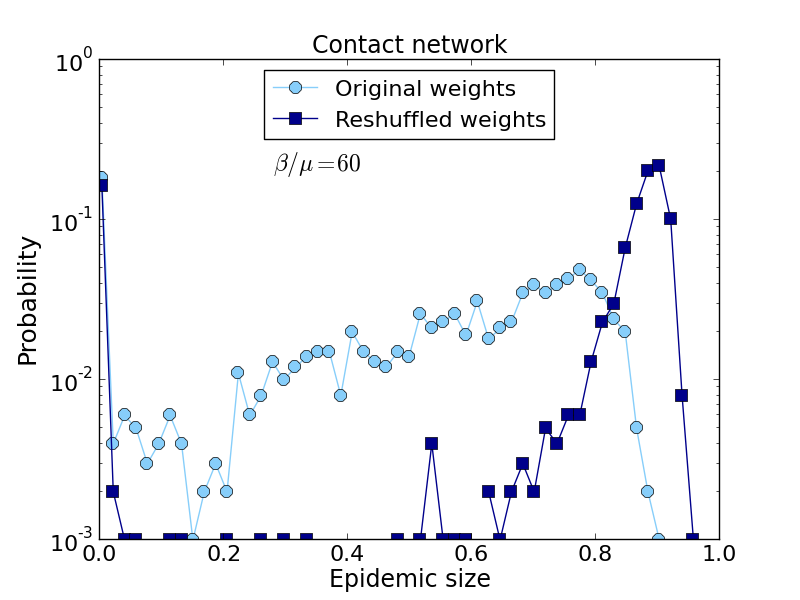}}
\subfigure[]
{\includegraphics[width=0.35\textwidth]{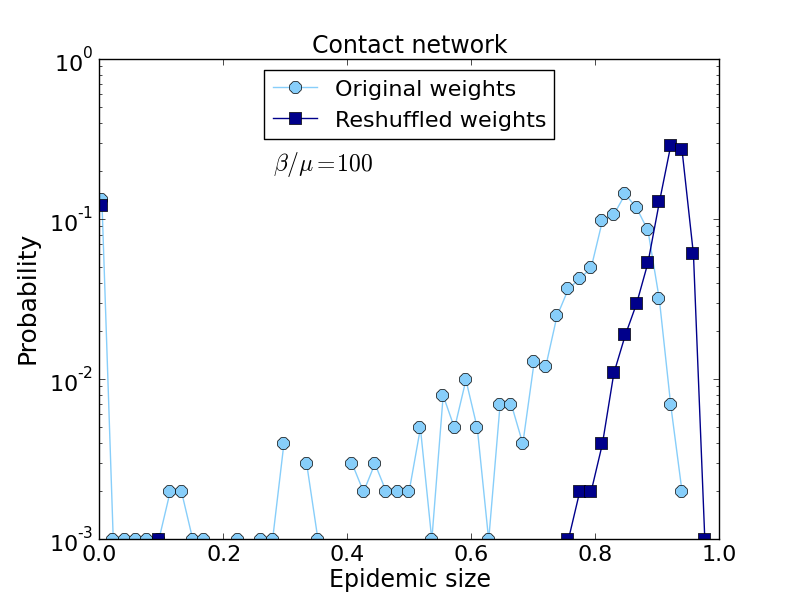}}
\caption{Comparison of the distributions of epidemic sizes of SIR spreading simulations performed on 
the whole contact network with ``Original weights'' and ``Reshuffled weights'' 
for two different values of  $\beta/\mu$.\label{fig:s2}}
\end{figure}

Figures \ref{fig:s1}-\ref{fig:s2} show however that simulations performed on the whole contact network with reshuffled
weights leads to a larger epidemic risk evaluation than when the original weights are used. This result carries on
to the case of RN, EGO and RE sampled networks: the use of randomly assigned weights systematically leads 
to larger epidemic sizes than the use of the real weights (Fig. \ref{fig:s3}). As the weight distribution is unchanged, this is the sign of the impact of some
correlations between weights and structure.

\begin{figure}[!ht]
\centering
\hspace{-10mm}
\subfigure[]
{\includegraphics[width=0.35\textwidth]{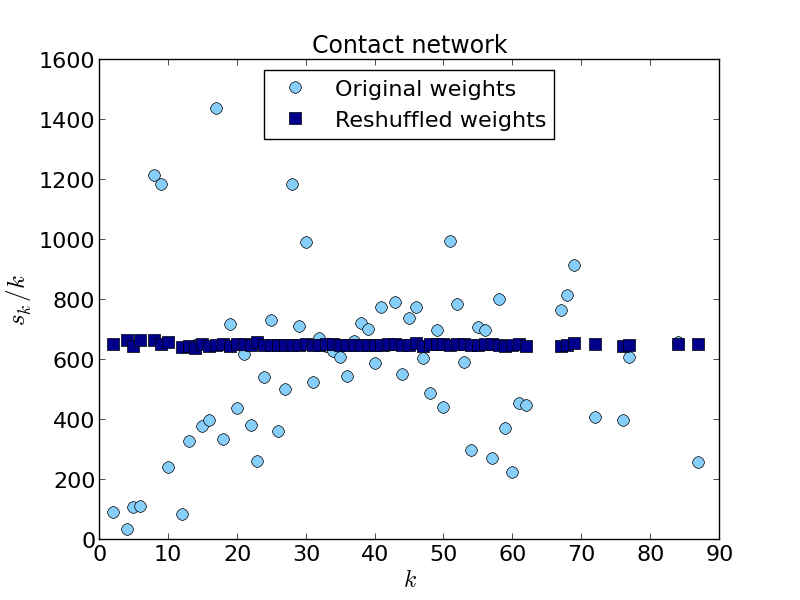}}
\subfigure[]
{\includegraphics[width=0.35\textwidth]{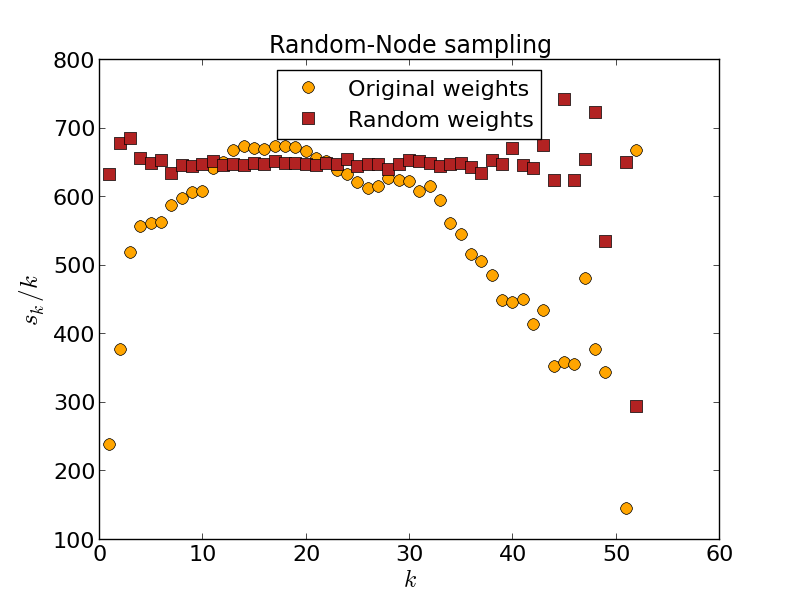}}\\
\hspace{-10mm}
\subfigure[]
{\includegraphics[width=0.35\textwidth]{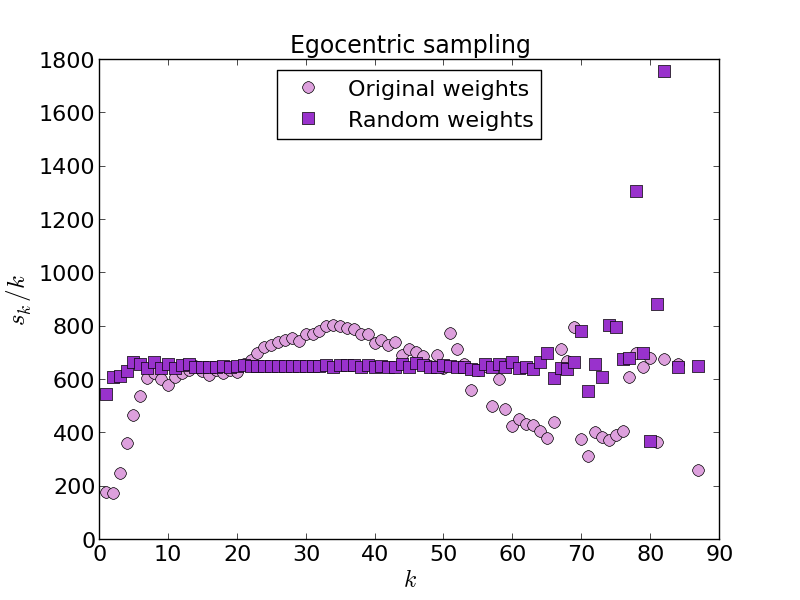}}
\subfigure[]
{\includegraphics[width=0.35\textwidth]{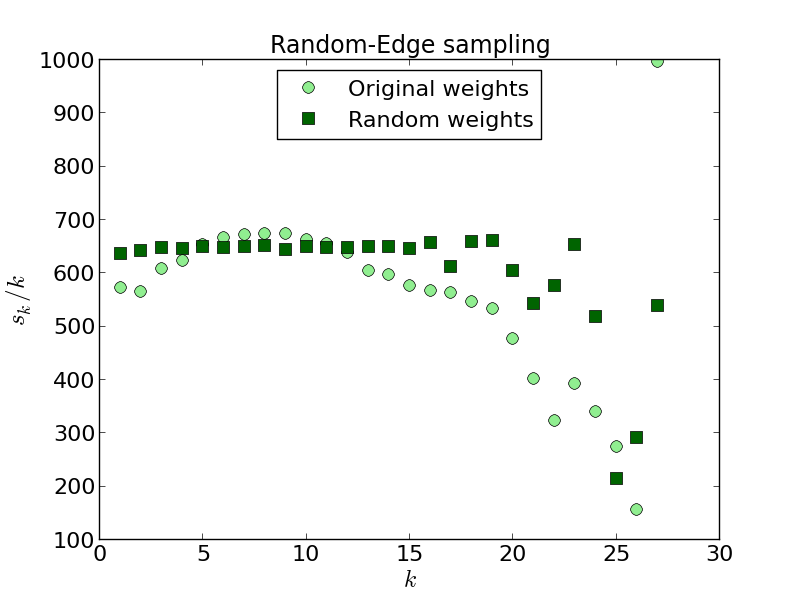}}
\caption{Comparison of average strength $s_k$
of nodes with degree $k$, divided by the degree $k$, as a function of $k$ between the ``Original weights'' and 
the ``Reshuffled/Random weights'' cases for ({a}) the contact network,
({b}) the RN-sampled network, ({c}) the EGO-sampled network, ({d}) the RE-sampled network.
\label{fig:s4}}
\end{figure}

\begin{figure}[!ht]
\centering
\hspace{-10mm}
\subfigure[]
{\includegraphics[width=0.35\textwidth]{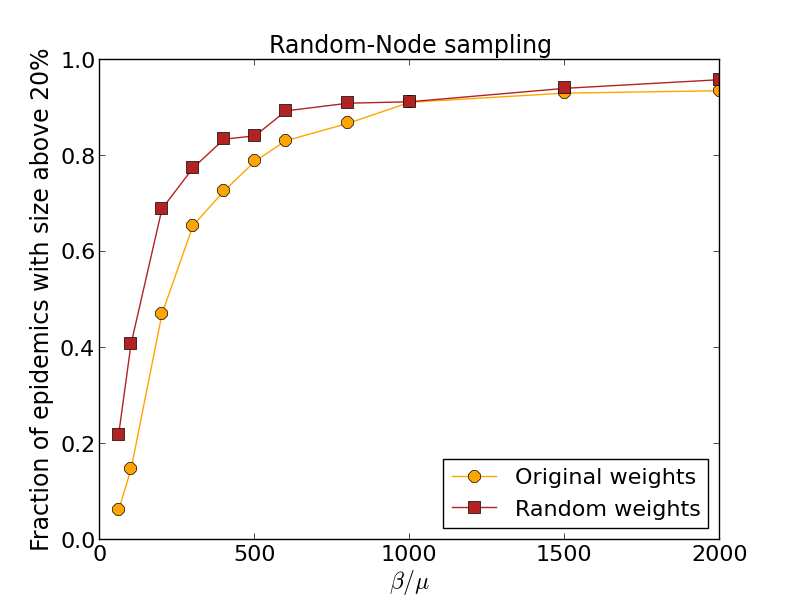}}
\subfigure[]
{\includegraphics[width=0.35\textwidth]{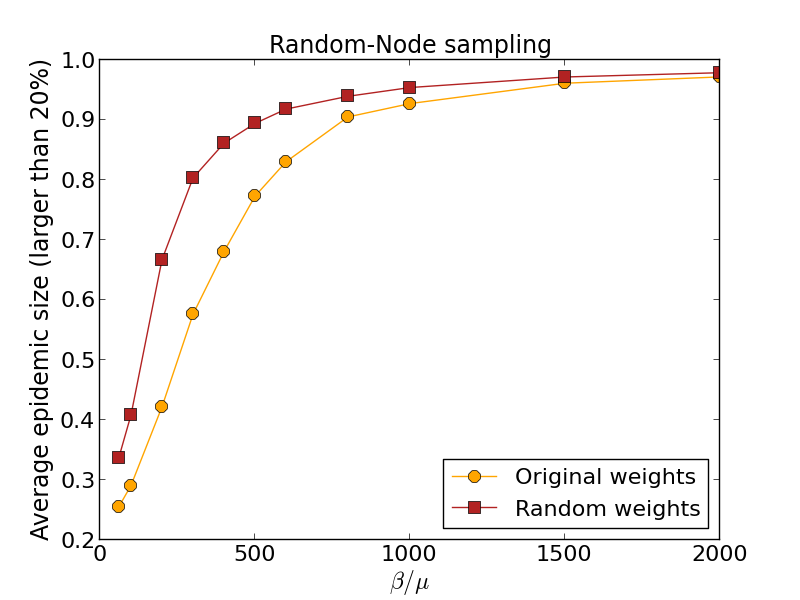}}
\subfigure[]
{\includegraphics[width=0.35\textwidth]{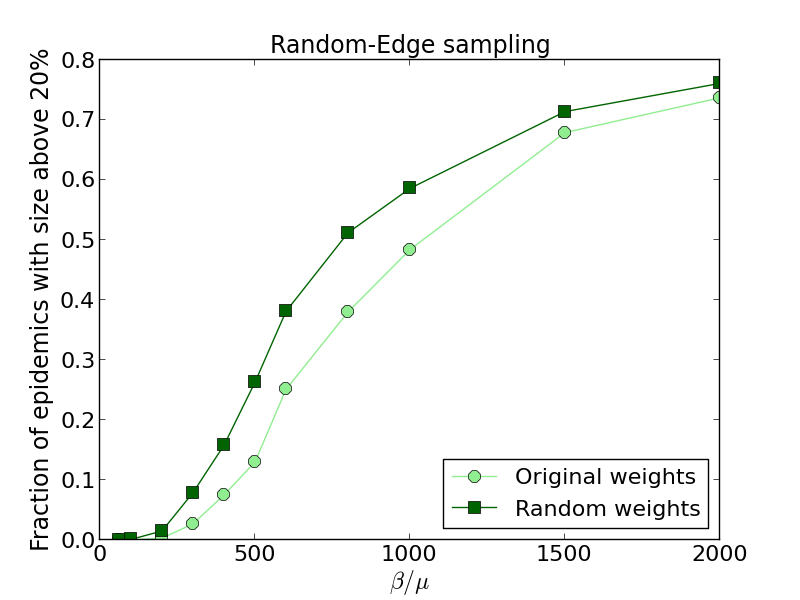}}
\subfigure[]
{\includegraphics[width=0.35\textwidth]{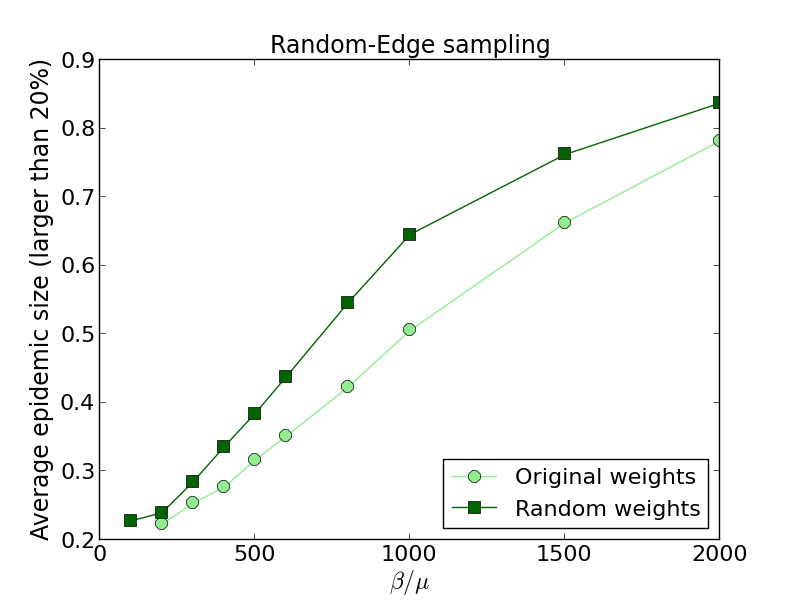}}
\subfigure[]
{\includegraphics[width=0.35\textwidth]{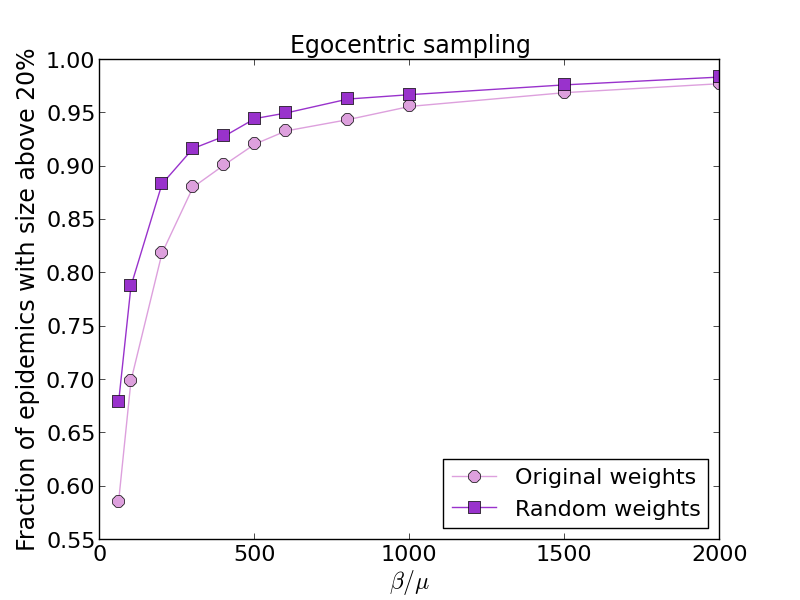}}
\subfigure[]
{\includegraphics[width=0.35\textwidth]{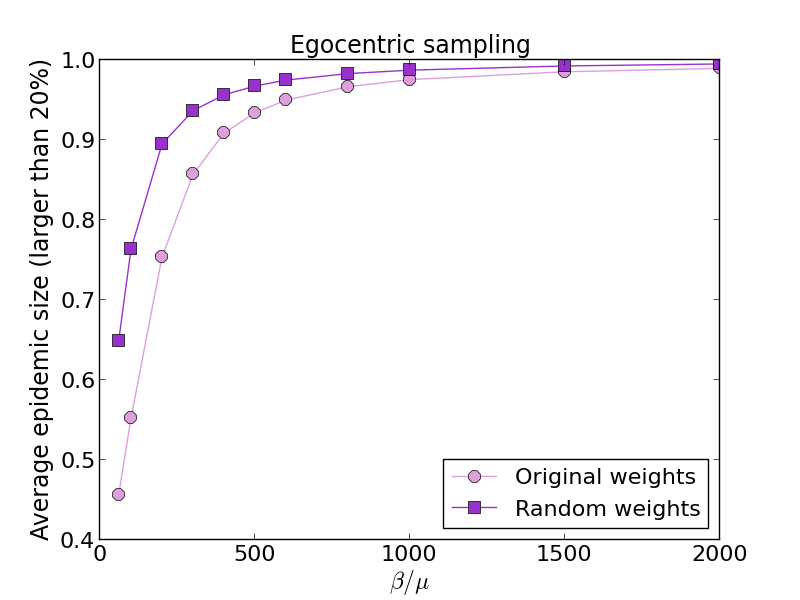}}
\caption{\textbf{Outcome of SIR spreading simulations performed on contact networks sampled
with the RN, RE and EGO methods, and with weights assigned either through the 
 ``Original weights'' or ``Random weights'' procedures.} 
(a),(c),(e) Fraction of epidemics with size above 20\% (at least 20\% of recovered 
individuals at the end of the SIR process) as a function of $\beta/\mu$. 
(b),(d),(f) Average size of epidemics with size larger than 20\% as a function of $\beta/\mu$.\label{fig:s3}}
\end{figure}

Figure \ref{fig:s4} shows that this is indeed the case: it displays the ratio 
$s_k/k$ as a function of $k$ for the contact network and the RN, EGO and RE sampled networks, where 
$s_k$ is the average strength of nodes of degree $k$. When weights are shuffled or assigned at random, 
$s_k/k$ is independent of $k$. For the original weights however, a distinct trend is observed, with smaller strengths
at large $k$ than for the reshuffled weights. As a consequence, the hubs have smaller spreading power than expected
by random chance, and the epidemic spread is hindered, leading to smaller epidemic risk.  

\subsection{Friendship network}

\begin{figure}[!ht]
\centering
\hspace{-10mm}
\subfigure[]
{\includegraphics[width=0.35\textwidth]{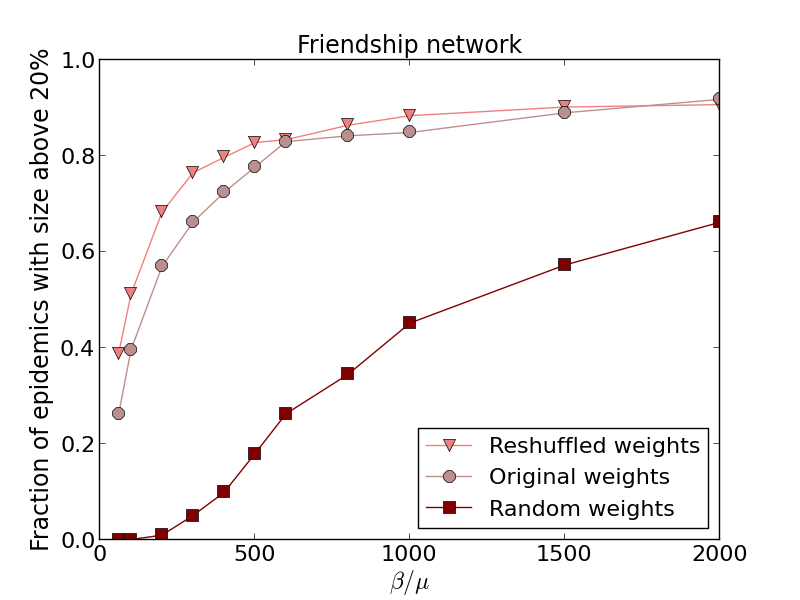}}
\subfigure[]
{\includegraphics[width=0.35\textwidth]{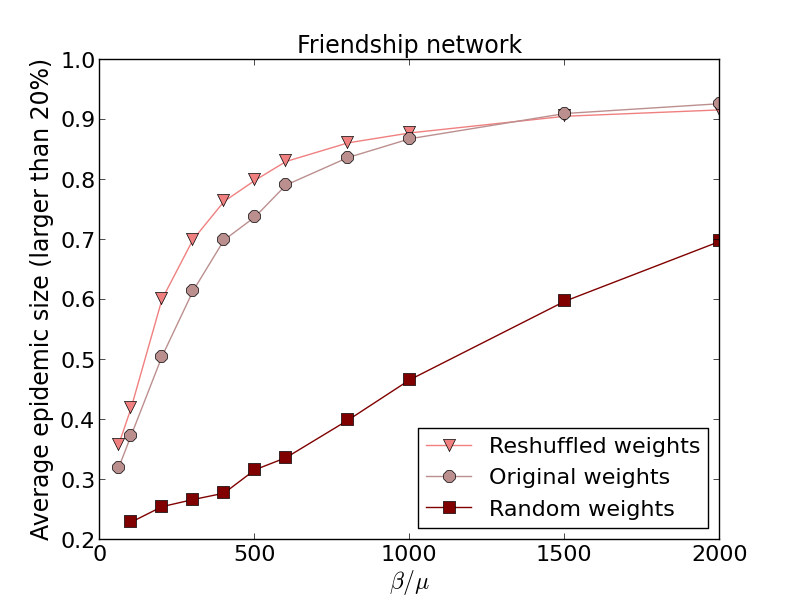}}
\caption{\textbf{Outcome of SIR spreading simulations performed on the friendship network with ``Original weights'', `
`Reshuffled weights'' and ``Random weights''.} 
({a}) Fraction of epidemics with size above 20\% (at least 20\% of recovered individuals at the end of the SIR process) as a function of  $\beta/\mu$. 
({b}) Average size of epidemic with size above 20\% as a function of $\beta/\mu$.\label{fig:s7}}
\end{figure}

\begin{figure}[!ht]
\centering
\hspace{-10mm}
\subfigure[]
{\includegraphics[width=0.35\textwidth]{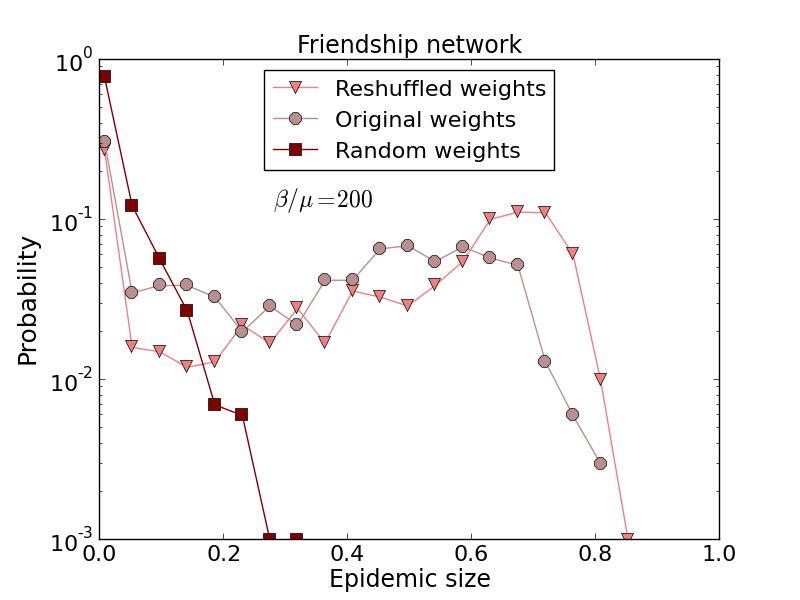}}
\subfigure[]
{\includegraphics[width=0.35\textwidth]{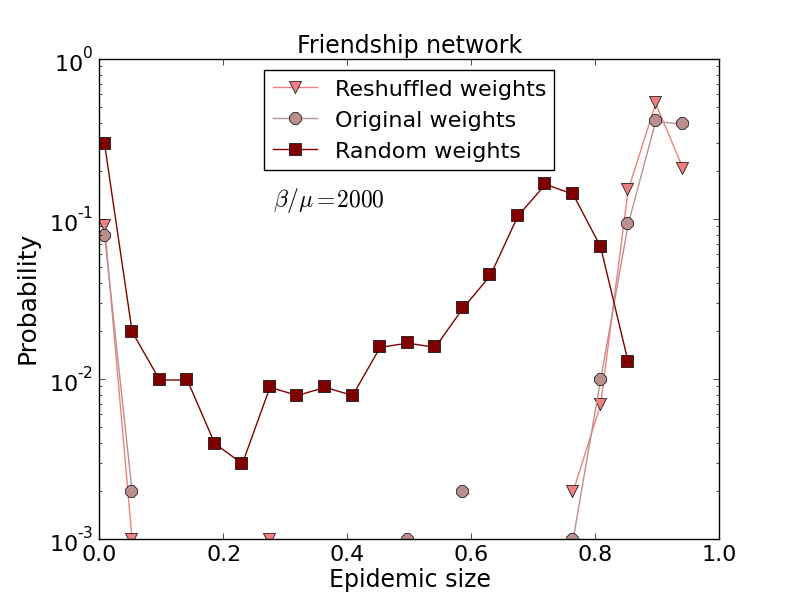}}
\caption{Comparison of the distributions of epidemic sizes of SIR spreading simulations performed on the 
friendship network with ``Original weights'', ``Reshuffled weights'' and ``Random weights'' for two different values of $\beta/\mu$.\label{fig:s8}}
\end{figure}

Figures \ref{fig:s7} and \ref{fig:s8} show
the outcomes of SIR simulations on the friendship network with weights assigned in the three
different ways described above. The size of epidemics is a little higher in the case of ``Reshuffled weights'' than in the case of ``Original weights'':
this is due to the same mechanism as for the contact network, i.e., to correlations between weight and structure that are destroyed
by the reshuffling. 

\begin{figure}[!ht]
\centering
{\includegraphics[width=0.5\textwidth]{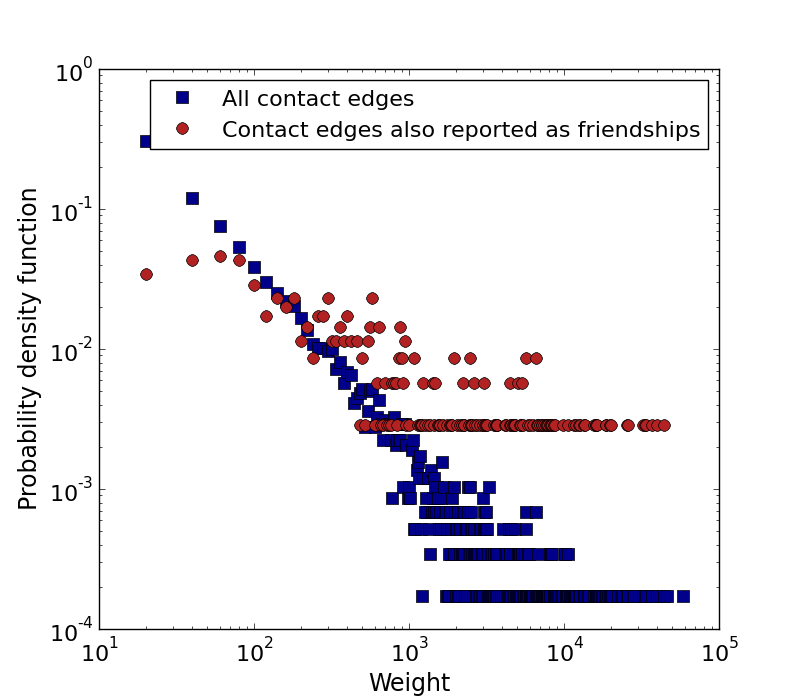}}
\caption{Distribution of weights for different kinds of edges 
in the contact network : (i) all edges, 
(ii) edges corresponding to a reported 
friendship (i.e., present in both friendship and contact networks). \label{fig:s9}}
\end{figure}

Simulations on the network with ``Random weights'' lead on the other hand to a much smaller epidemic risk. 
As shown in Figure \ref{fig:s9} and discussed in \cite{Mastrandrea:2015_2} indeed, the friendship links that are also
present in the contact network tend to correspond to larger cumulative contact durations: the distribution
of weights of the links present in both networks is not the same as the overall distribution of cumulative contact durations.
Since the latter is used in the ``Random weights'' assignment procedure, the average weight is larger in the 
``Original weights'' case, and this of course favours the spread.

\subsection{EGOref sampling procedure}

\begin{figure}[!ht]
\centering
\hspace{-10mm}
\subfigure[]
{\includegraphics[width=0.35\textwidth]{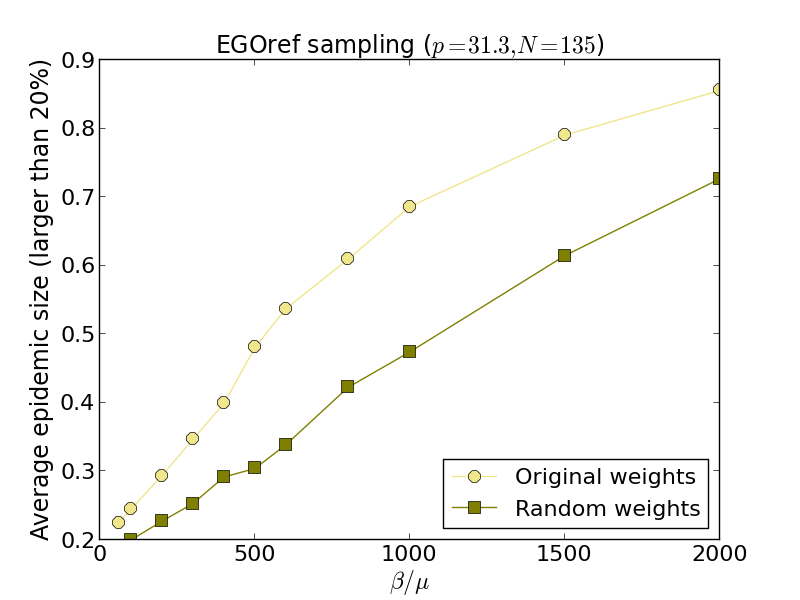}}
\subfigure[]
{\includegraphics[width=0.35\textwidth]{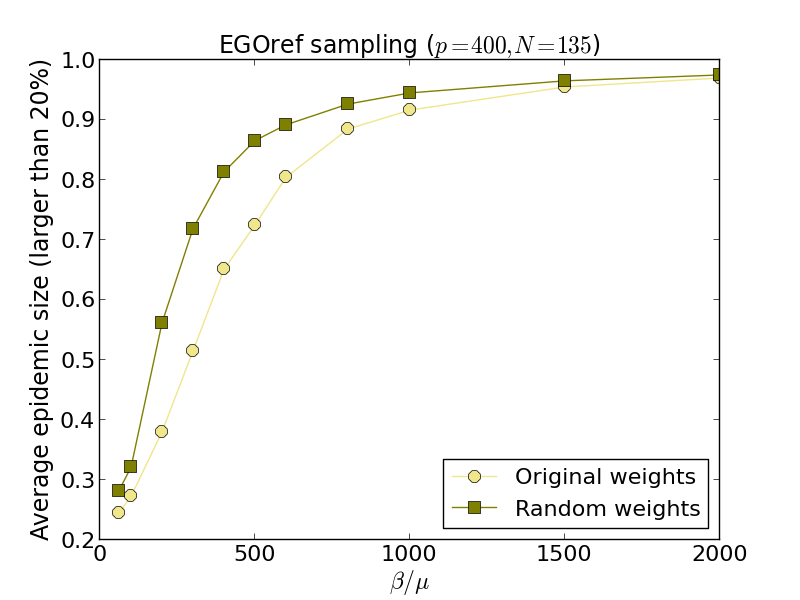}}\\
\hspace{-10mm}
\subfigure[]
{\includegraphics[width=0.35\textwidth]{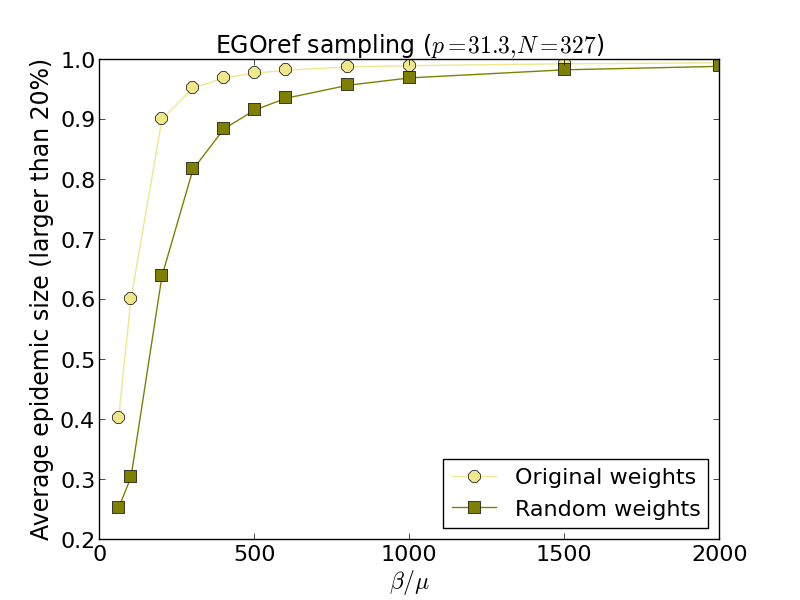}}
\subfigure[]
{\includegraphics[width=0.35\textwidth]{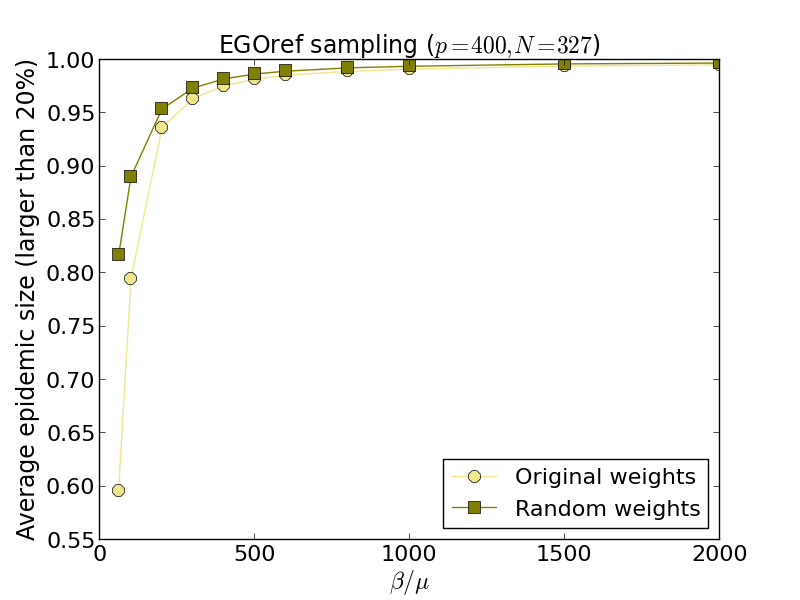}}
\caption{\textbf{Average size of epidemic (with size above 20\%) for SIR spreading 
simulations performed on EGOref-sampled network with ``Original weights'' and ``Random weights'' for 
four different couples of $p$ and $N$ : } (\textbf{a}) $p=31.3, N=135$, (\textbf{b}) $p=400, N=135$, (\textbf{c}) 
$p=31.3, N=327$, (\textbf{d}) $p=400, N=327$.\label{fig:s10}}
\end{figure}

The case of the EGOref sampling procedure is interesting as it combines the two effects discussed above, and 
the difference between the outcomes of simulations performed on networks using the ``Original weights'' and ``Random weights''
assignment procedures depends on the parameters of the sampling procedure $p$ and $N$, as shown in Figure \ref{fig:s10}. 
For small $p$, the average epidemic size is higher in the case of ``Original weights'' whereas for large $p$, 
it is higher in the ``Random weights'' case. This can be explained by the two following competing
effects:
\begin{itemize}
 \item at small $p$, relatively few edges are selected in the contact network, and each ego selects
 preferentially links with large weights. The resulting distribution of original weights is thus biased towards large weights,
 and the weights are on average larger than when using weights taken at random from the overall distribution
 of weights of the contact network. This tends to favour the spread and thus leads to a larger epidemic risk for
 ``Original weights'' than for ``Random weights''.
 
 \item at large $p$, the probability to select an edge is large even for links with small weights.
As a result, the distribution of weights of sampled links becomes close to the global distribution of weights in the contact network.
The correlations between weights and structure present in the contact network can then play a role and act in the same way as
for the RN, RE and EGO sampling methods: a random assignment of weights destroys the correlations and favours the spread.
 \end{itemize}
We finally note that the value of $N$ does not change the sign of the difference between the epidemic risk obtained by the
two weight assignment procedures, but only its amplitude.

\section{Impact of the EGOref sampling on other data sets}

\begin{figure}[!ht]
\centering
\hspace{-10mm}
\subfigure[]
{\includegraphics[width=0.35\textwidth]{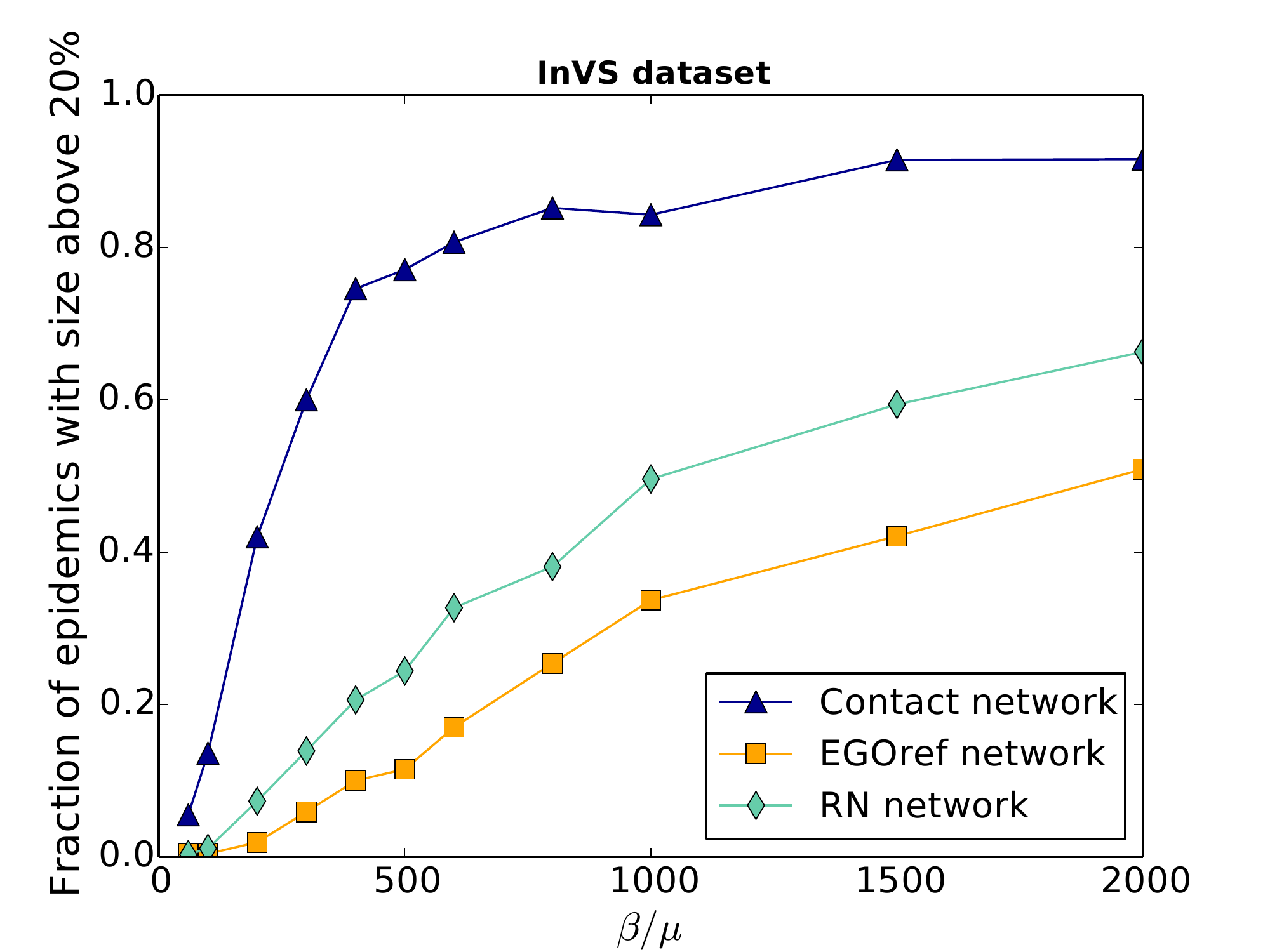}}
\subfigure[]
{\includegraphics[width=0.35\textwidth]{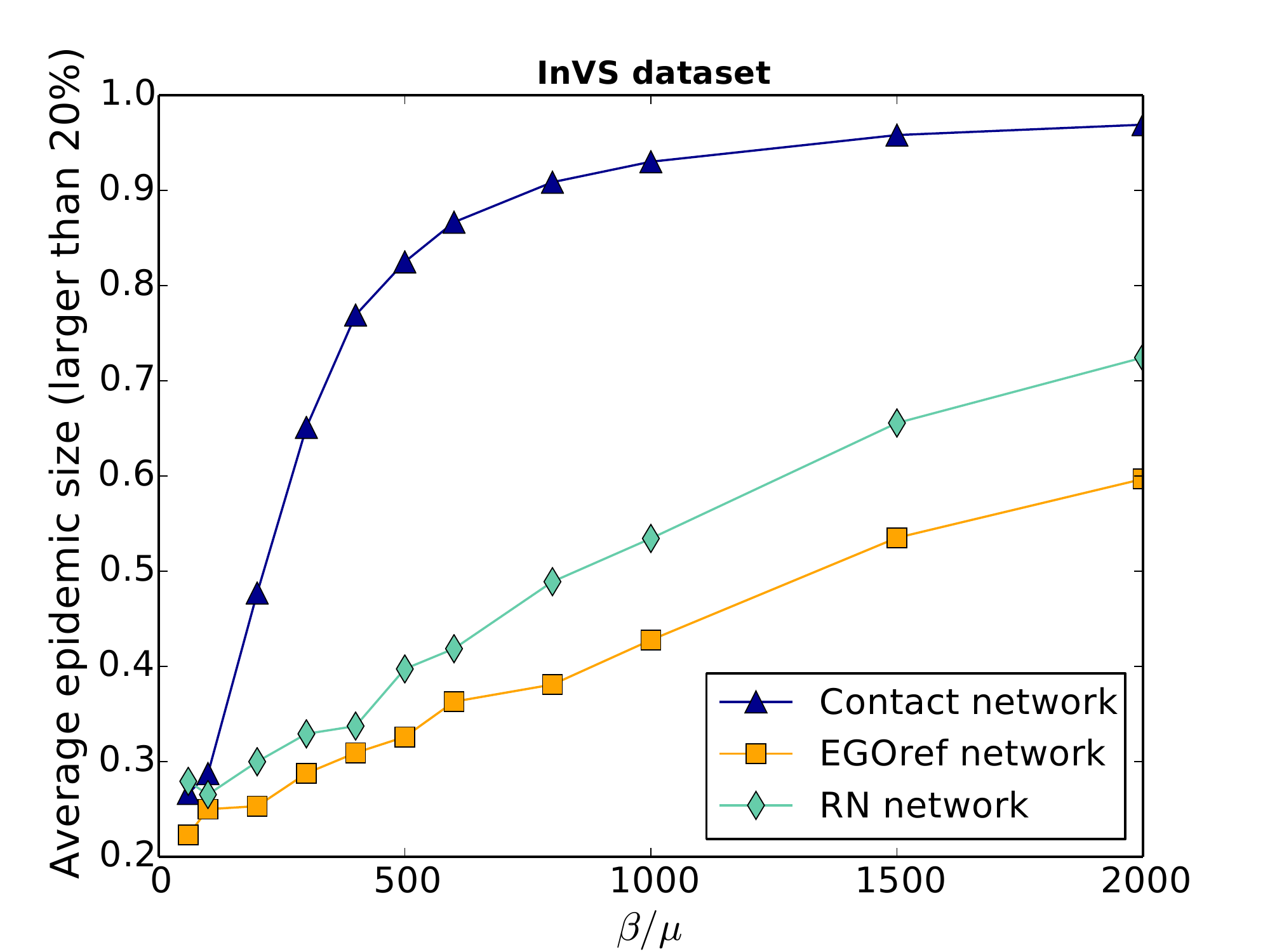}}
\caption{\textbf{Outcome of SIR spreading simulations performed on empirical and sampled contact networks (InVS data set).} 
We compare here the simulations on the original contact network with a sampled network using the EGOref sampling procedure 
with $p=31.3$ and $N=165$ nodes (corresponding to a sampling fraction equal to the case of the main text) and with
the Random Nodes case (still with $N=165$ nodes).
({a}) Fraction of epidemics with size above 20\% (at least 20\% of recovered individuals at the end of the SIR process) as a function of 
$\beta/\mu$. ({b}) Average size of epidemics with size above 20\% as a function of $\beta/\mu$. \label{fig:s11}}
\end{figure}

\begin{figure}[!ht]
\centering
\hspace{-10mm}
\subfigure[]
{\includegraphics[width=0.35\textwidth]{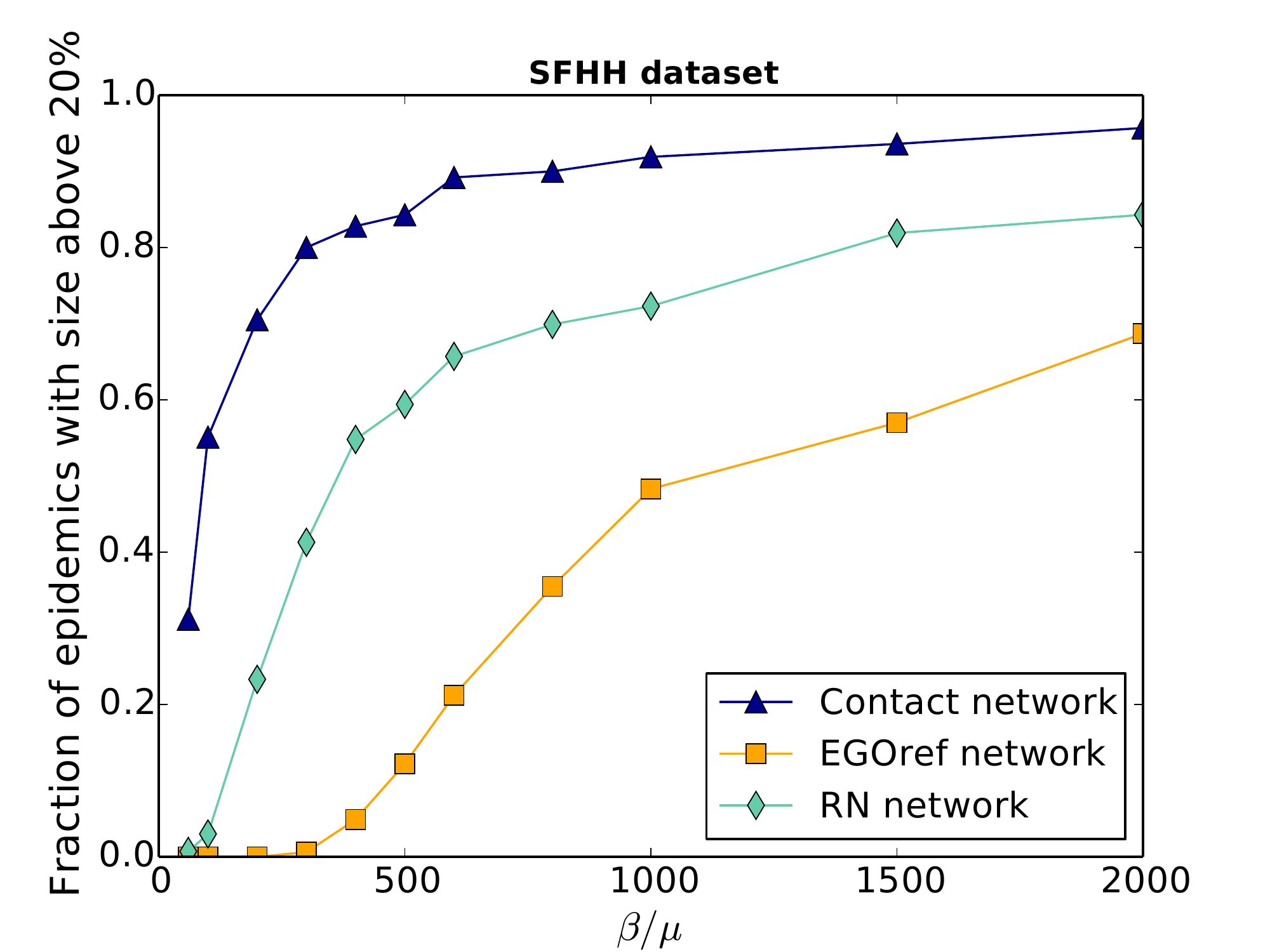}}
\subfigure[]
{\includegraphics[width=0.35\textwidth]{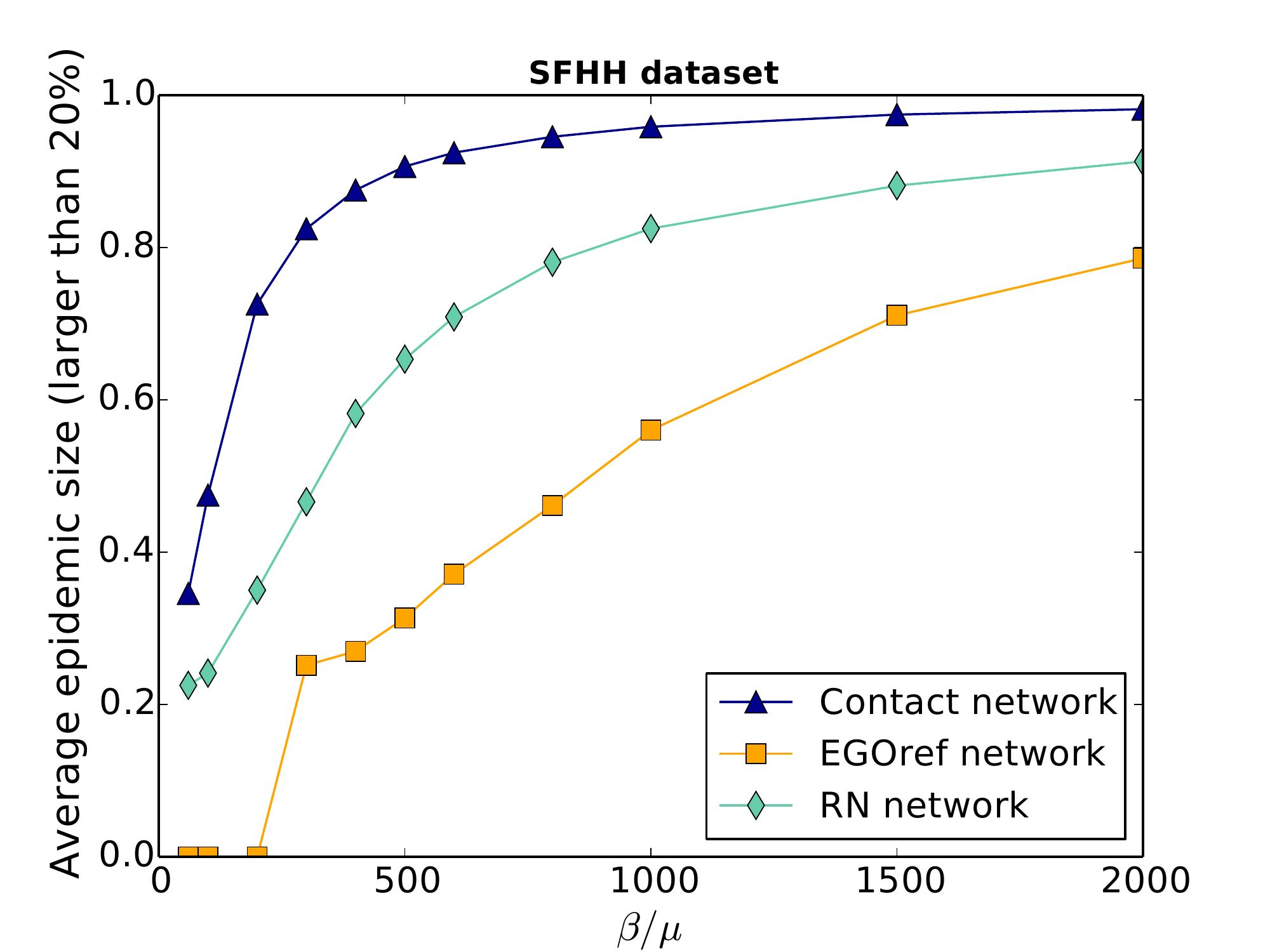}}
\caption{\textbf{Outcome of SIR spreading simulations performed on empirical and sampled contact networks (SFHH data set).} 
We compare here the simulations on the original contact network with a sampled network using the EGOref sampling procedure 
with $p=31.3$ and $N=37$ nodes (corresponding to a sampling fraction equal to the case of the main text) and with
the Random Nodes case (still with $N=37$ nodes).
({a}) Fraction of epidemics with size above 20\% (at least 20\% of recovered individuals at the end of the SIR process) as a function of 
$\beta/\mu$. ({b}) Average size of epidemics with size above 20\% as a function of $\beta/\mu$. \label{fig:s12}}
\end{figure}

\begin{figure}[!ht]
\centering
\hspace{-10mm}
\subfigure[$\beta/\mu = 500$]
{\includegraphics[width=0.35\textwidth]{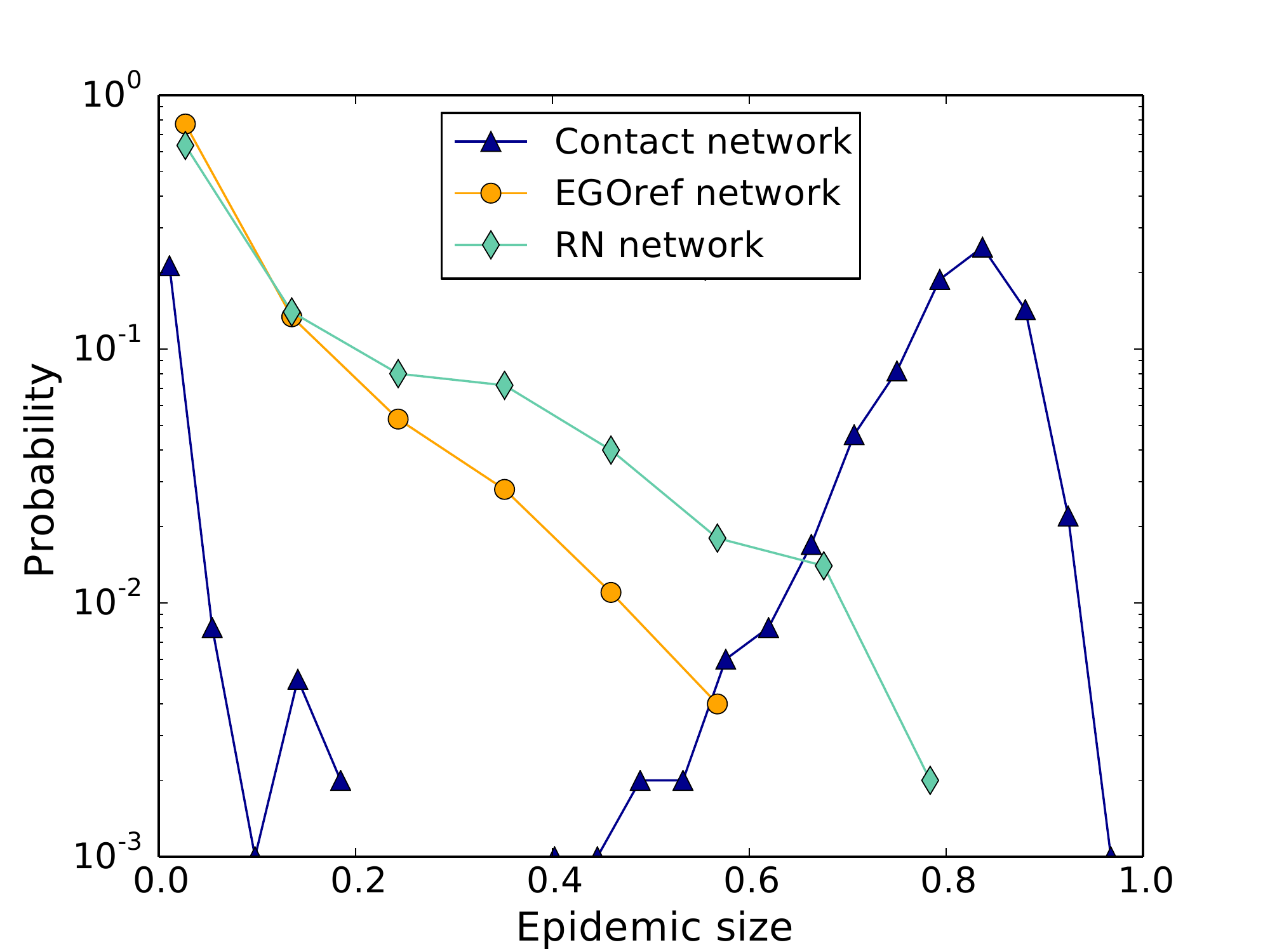}}
\subfigure[$\beta/\mu = 1000$]
{\includegraphics[width=0.35\textwidth]{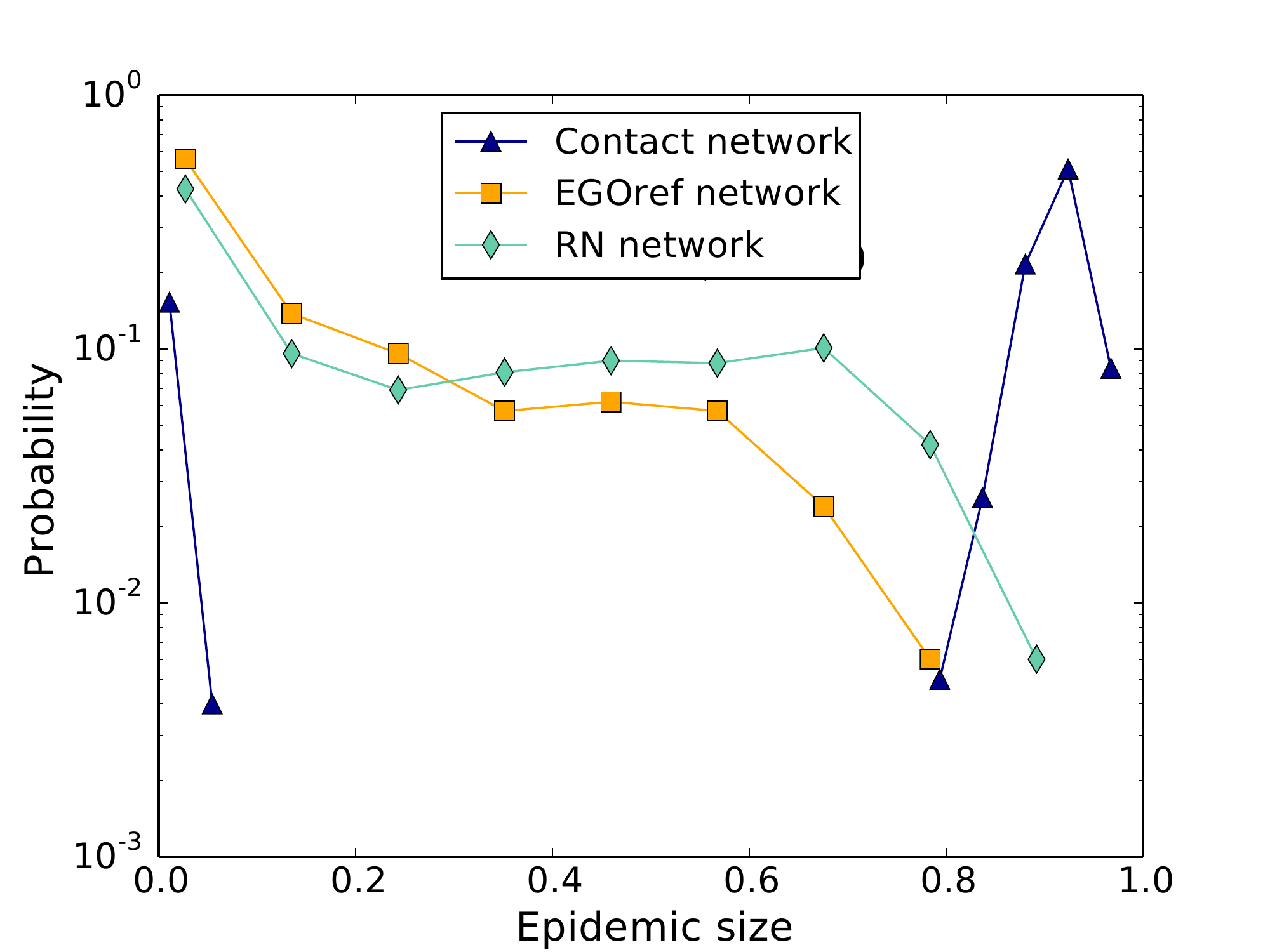}}\\
\hspace{-10mm}
\subfigure[$\beta/\mu = 1500$]
{\includegraphics[width=0.35\textwidth]{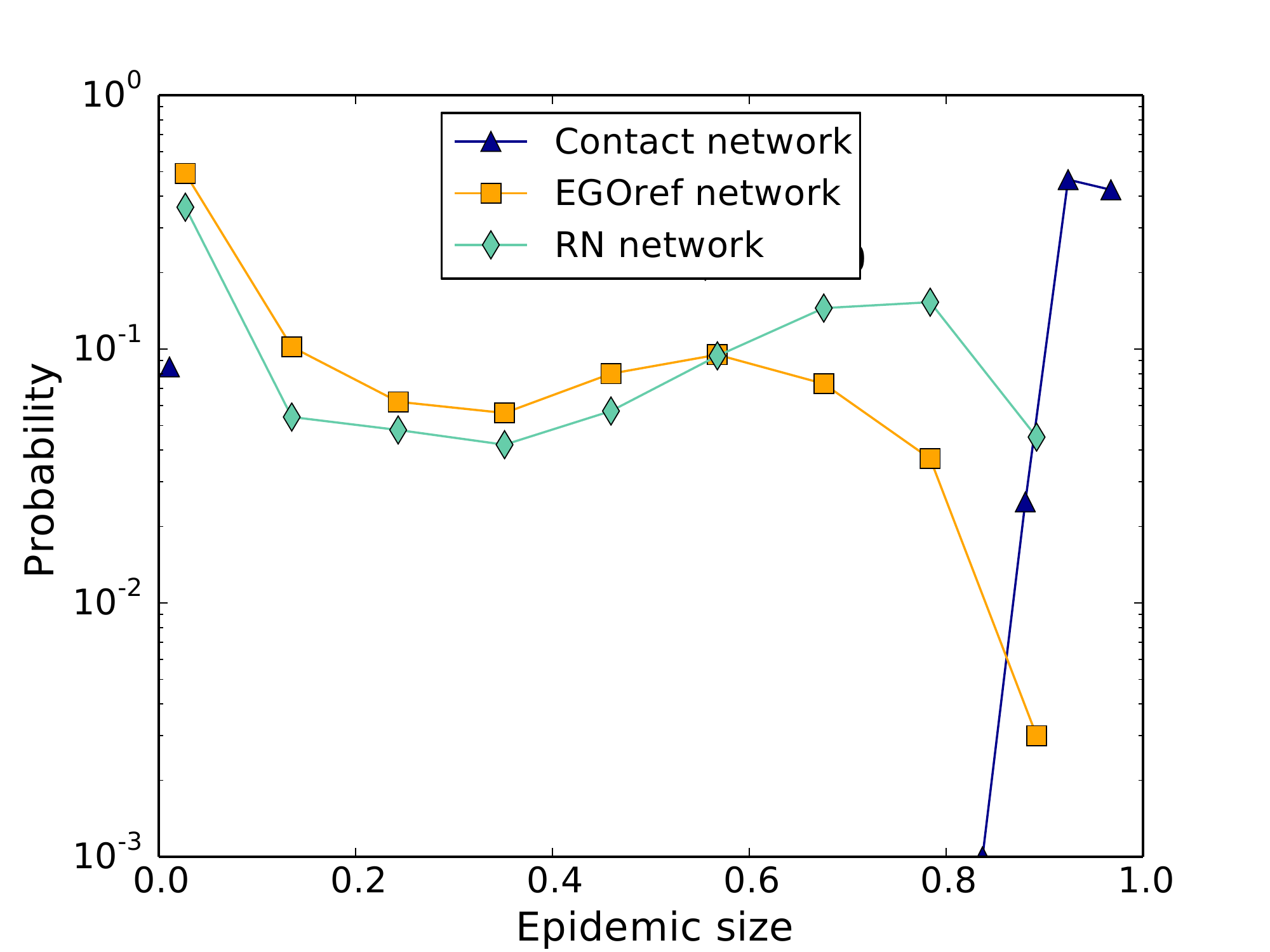}}
\subfigure[$\beta/\mu = 2000$]
{\includegraphics[width=0.35\textwidth]{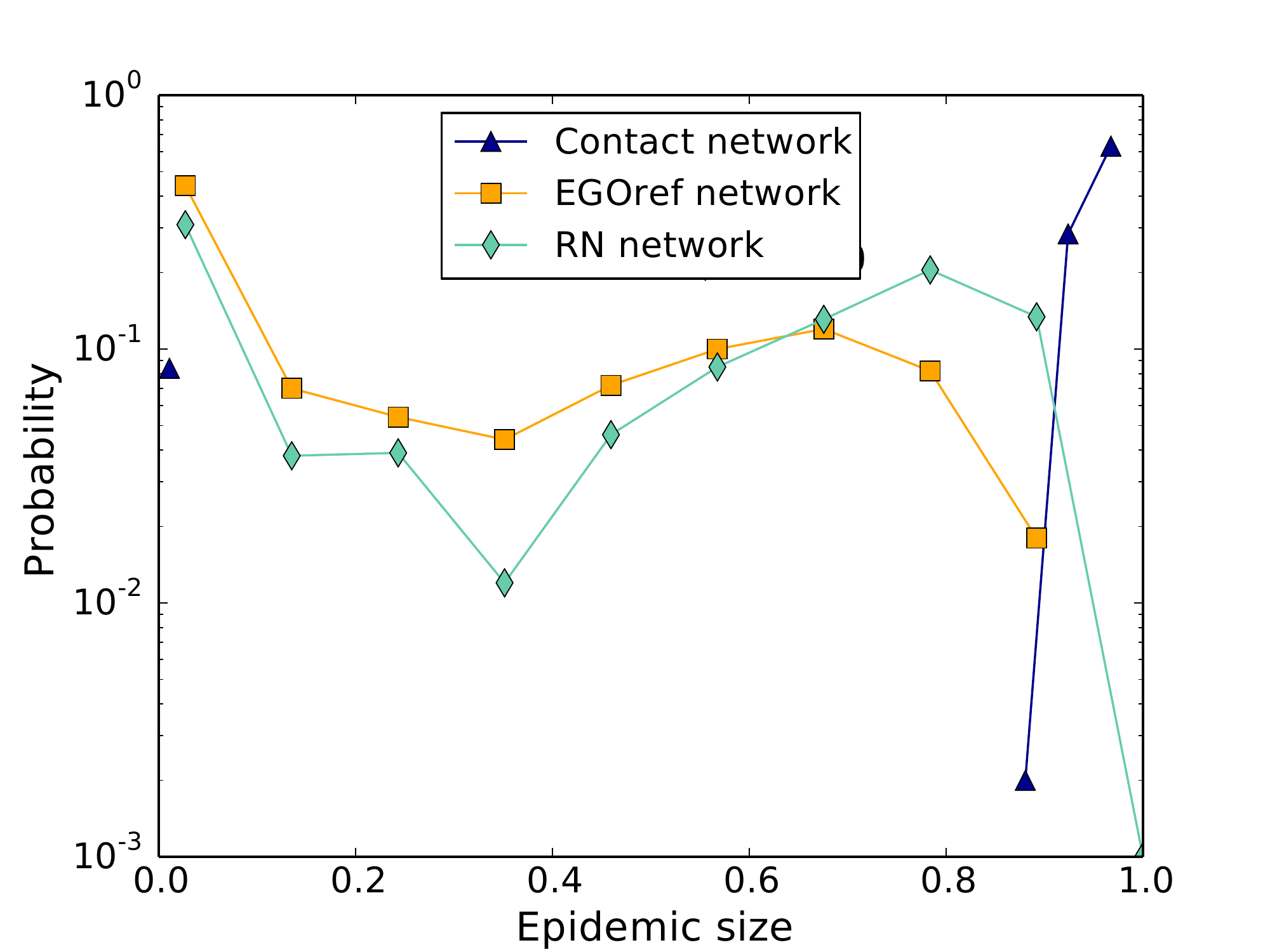}}
\caption{\textbf{Distributions of epidemic sizes of SIR spreading simulations (InVS data set).} We compare the distributions
of epidemic sizes for simulations performed on the original contact network and on the sampled  network using the EGOref sampling procedure 
with $p=31.3$ and $N=37$ nodes (corresponding to a sampling fraction equal to the case of the main text) and with
the Random Nodes case (still with $N=37$ nodes), for different values of $\beta/\mu$.
\label{fig:s13}}
\end{figure}

\begin{figure}[!ht]
\centering
\hspace{-10mm}
\subfigure[$\beta/\mu = 500$]
{\includegraphics[width=0.35\textwidth]{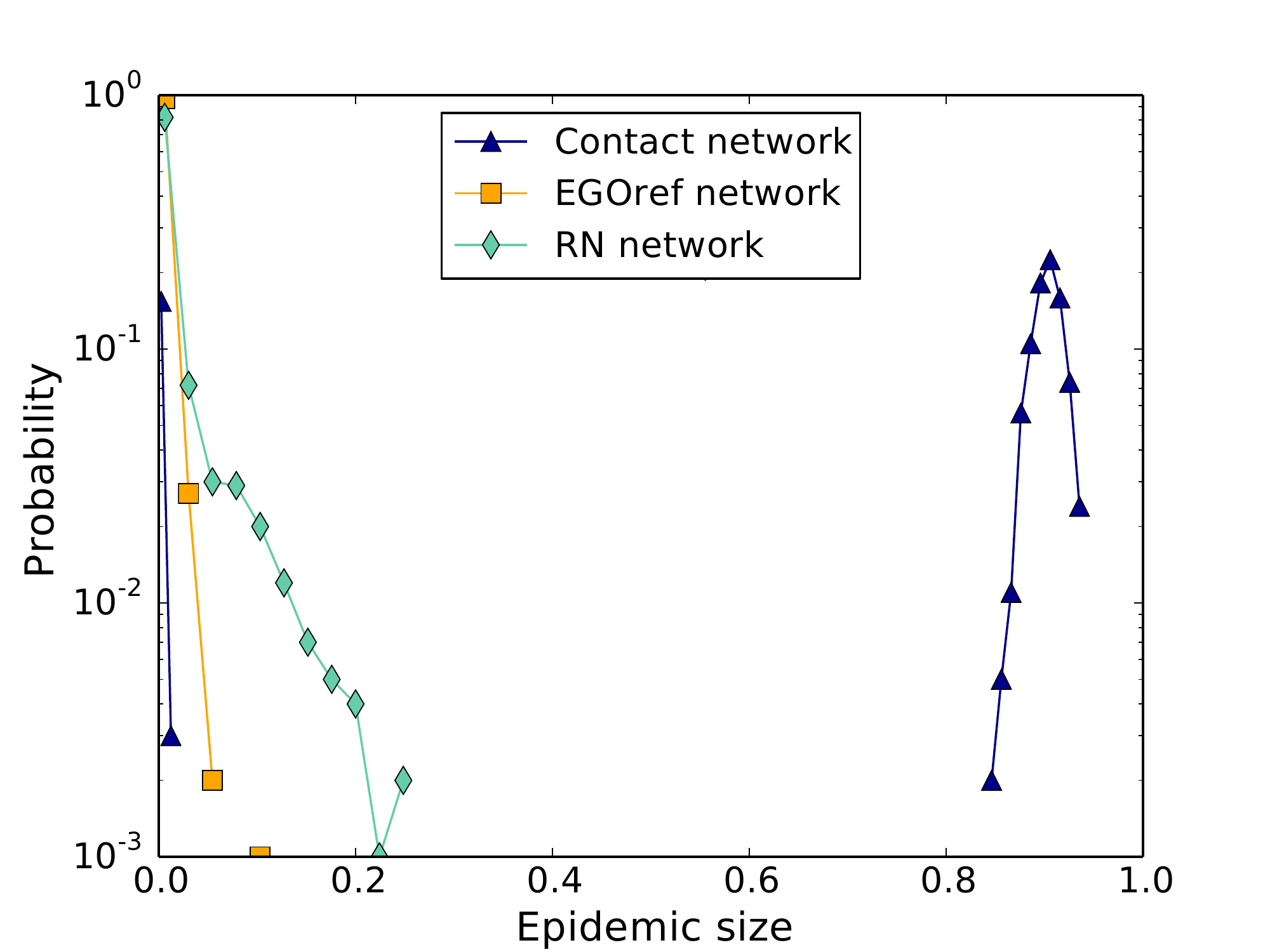}}
\subfigure[$\beta/\mu = 1000$]
{\includegraphics[width=0.35\textwidth]{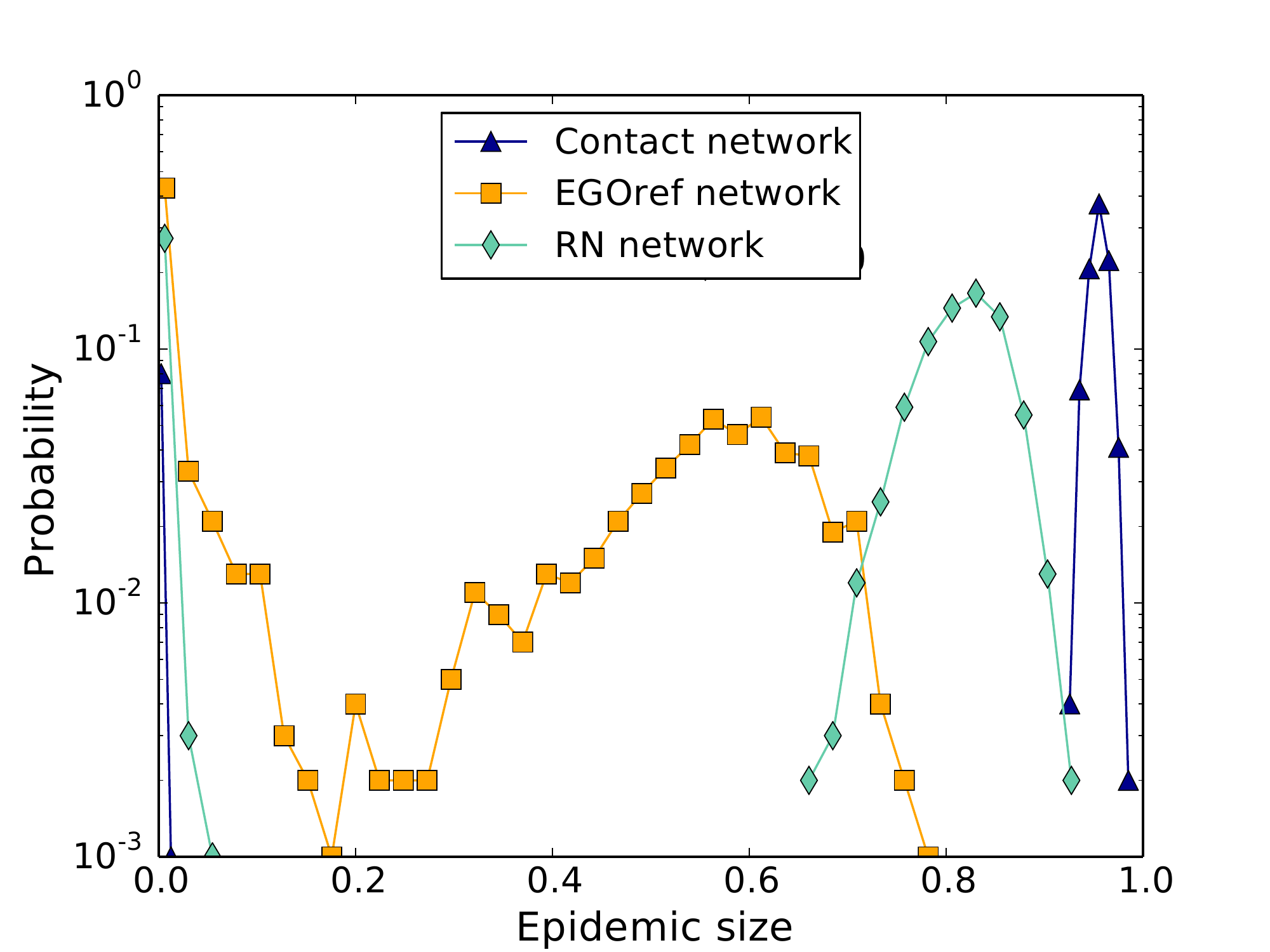}}\\
\hspace{-10mm}
\subfigure[$\beta/\mu = 1500$]
{\includegraphics[width=0.35\textwidth]{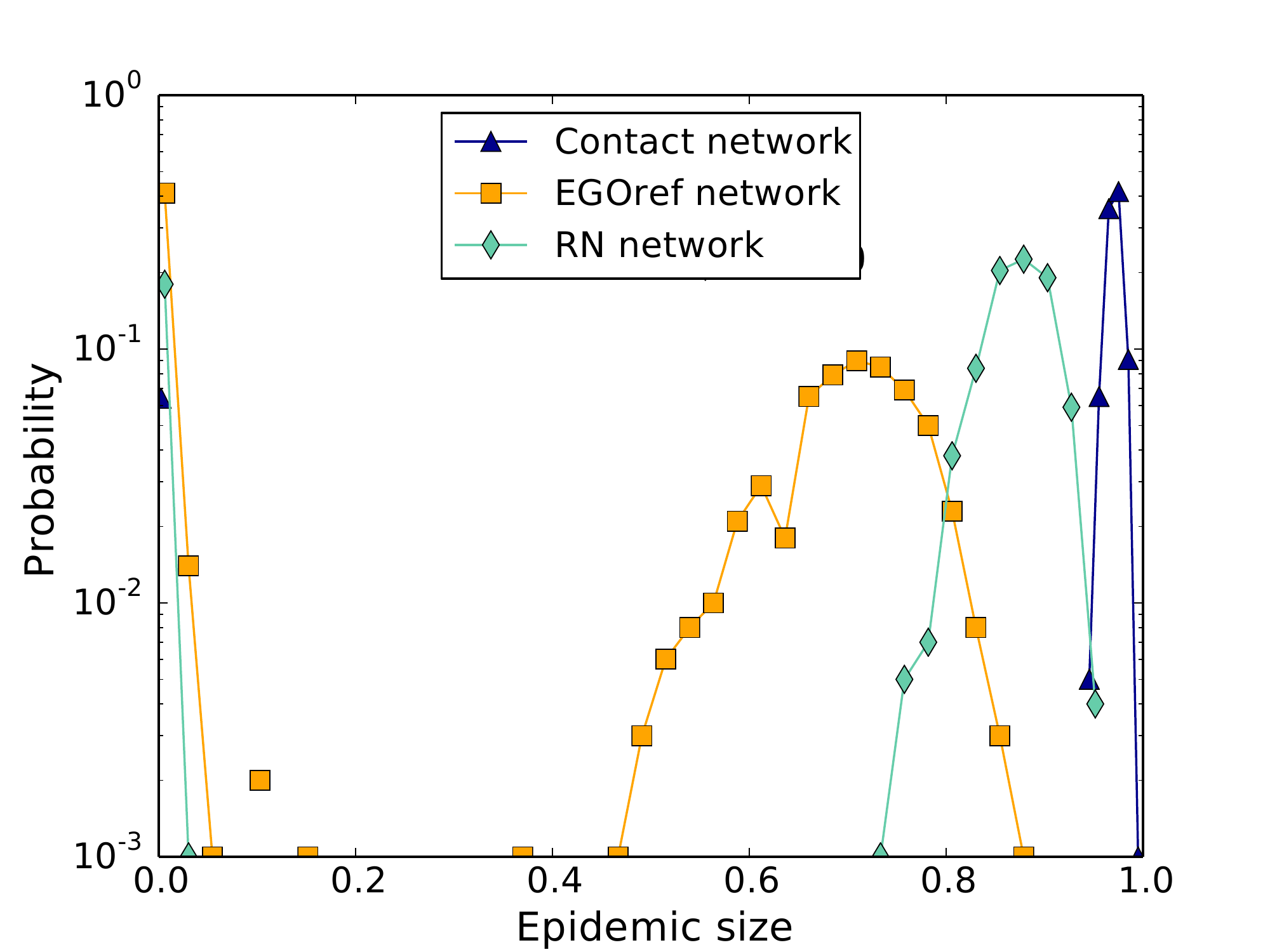}}
\subfigure[$\beta/\mu = 2000$]
{\includegraphics[width=0.35\textwidth]{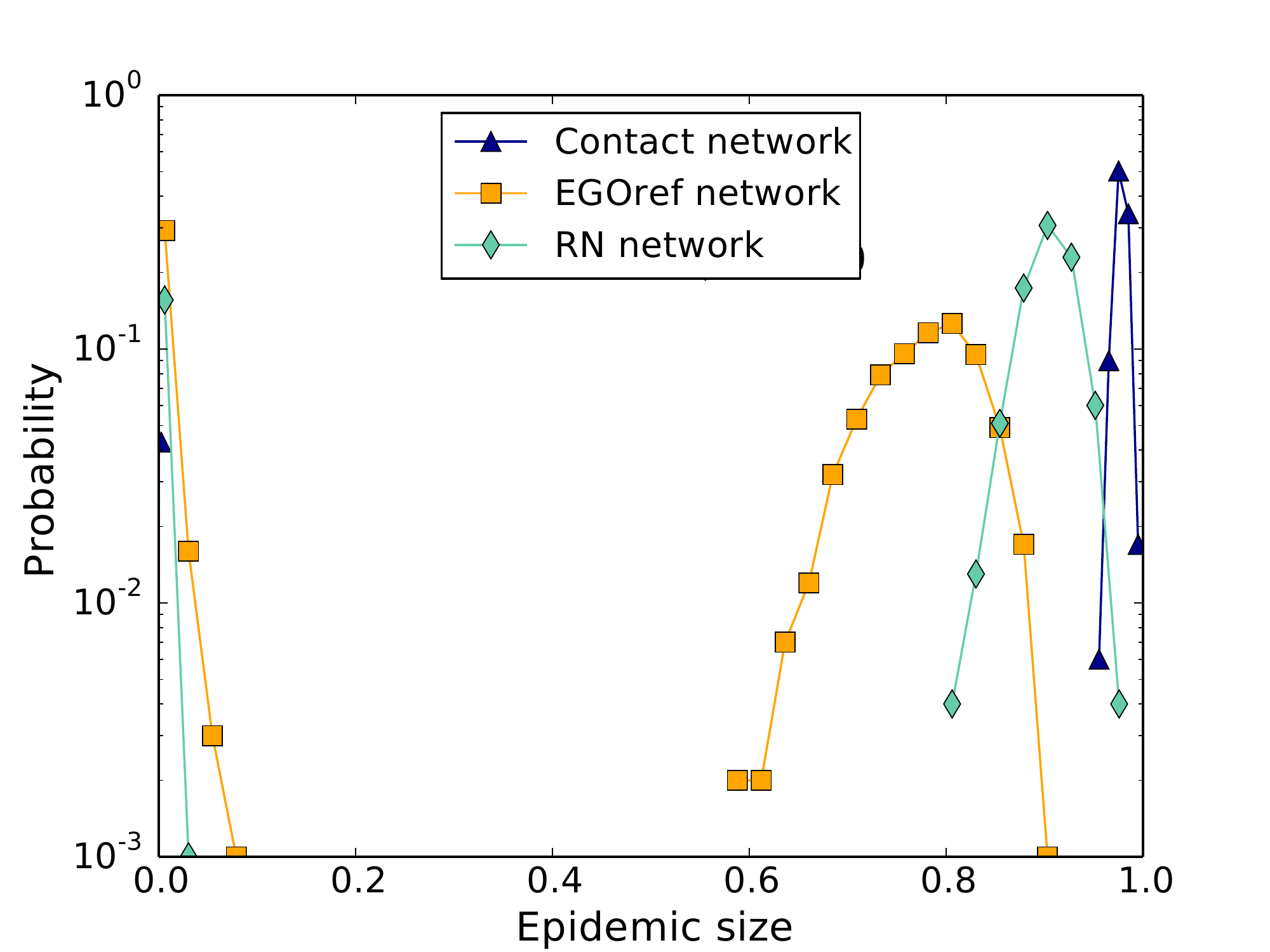}}
\caption{\textbf{Distributions of epidemic sizes of SIR spreading simulations (SFHH data set).} We compare the distributions
of epidemic sizes for simulations performed on the original contact network and on the sampled  network using the EGOref sampling procedure 
with $p=31.3$ and $N=165$ nodes (corresponding to a sampling fraction equal to the case of the main text) and with
the Random Nodes case (still with $N=165$ nodes), for different values of $\beta/\mu$.
\label{fig:s14}}
\end{figure}

We consider here two data sets describing contacts in (i) a conference (SFHH) and (ii) offices (InVS). 
Both data sets were described for instance in \cite{Genois:2015_2}. The SFHH data describes contacts between $403$
individuals during the two days of a conference, while the InVS data contains the contacts measured in offices during 
two weeks for $92$ individuals.
We first show in Fig.s \ref{fig:s11}-\ref{fig:s14} the case of the same parameters as in 
the main text, namely the same value of $p$ and the same fraction of
nodes, i.e. $41\%$. We also show in these figures the outcome of simulations performed on a population sampling
of the contact network with the same sampling fraction (RN sampling), in order to show separately the effects of population
sampling (which keeps the density of the contact network fixed \cite{Genois:2015_2}) and of the selection of the links with
probability proportional to their weights (EGOref mechanism). 

\begin{figure}[!ht]
\centering
\hspace{-10mm}
\subfigure[]
{\includegraphics[width=0.35\textwidth]{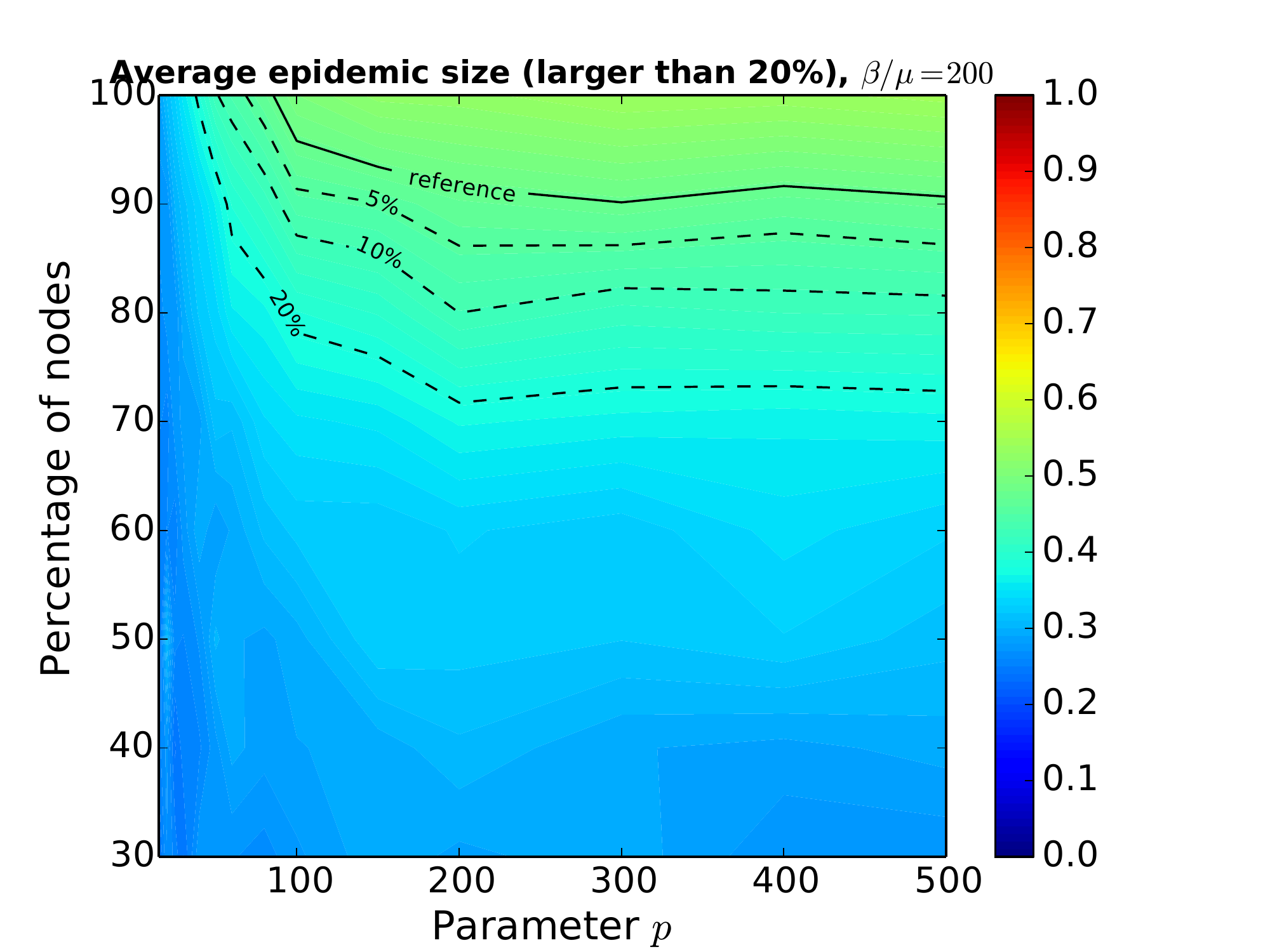}}
\subfigure[]
{\includegraphics[width=0.35\textwidth]{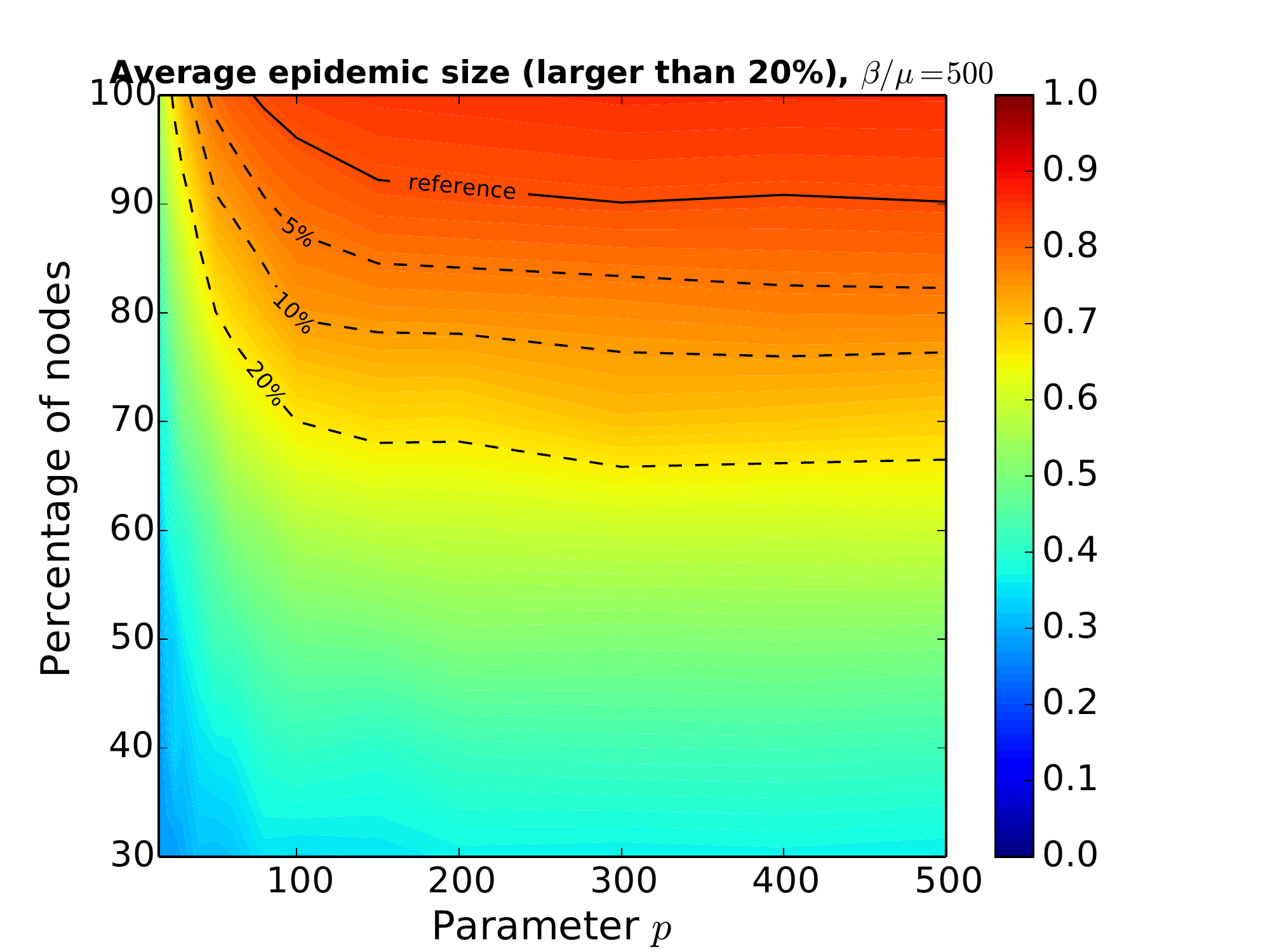}}\\
\caption{\textbf{Color maps of the average epidemic size for epidemics with size above 20\% for several
values of $\beta/\mu$ (InVS data set).}
When no epidemics has size above 20\%, the value is zero. The three dashed lines represent the value of
average epidemic size at 5\%, 10\%, 20\% of the reference value (solid line),
which corresponds to the average epidemic size of the SIR spreading simulations performed on the contact network.\label{fig:s15}}
\end{figure}

\begin{figure}[!ht]
\centering
\hspace{-10mm}
\subfigure[]
{\includegraphics[width=0.5\textwidth]{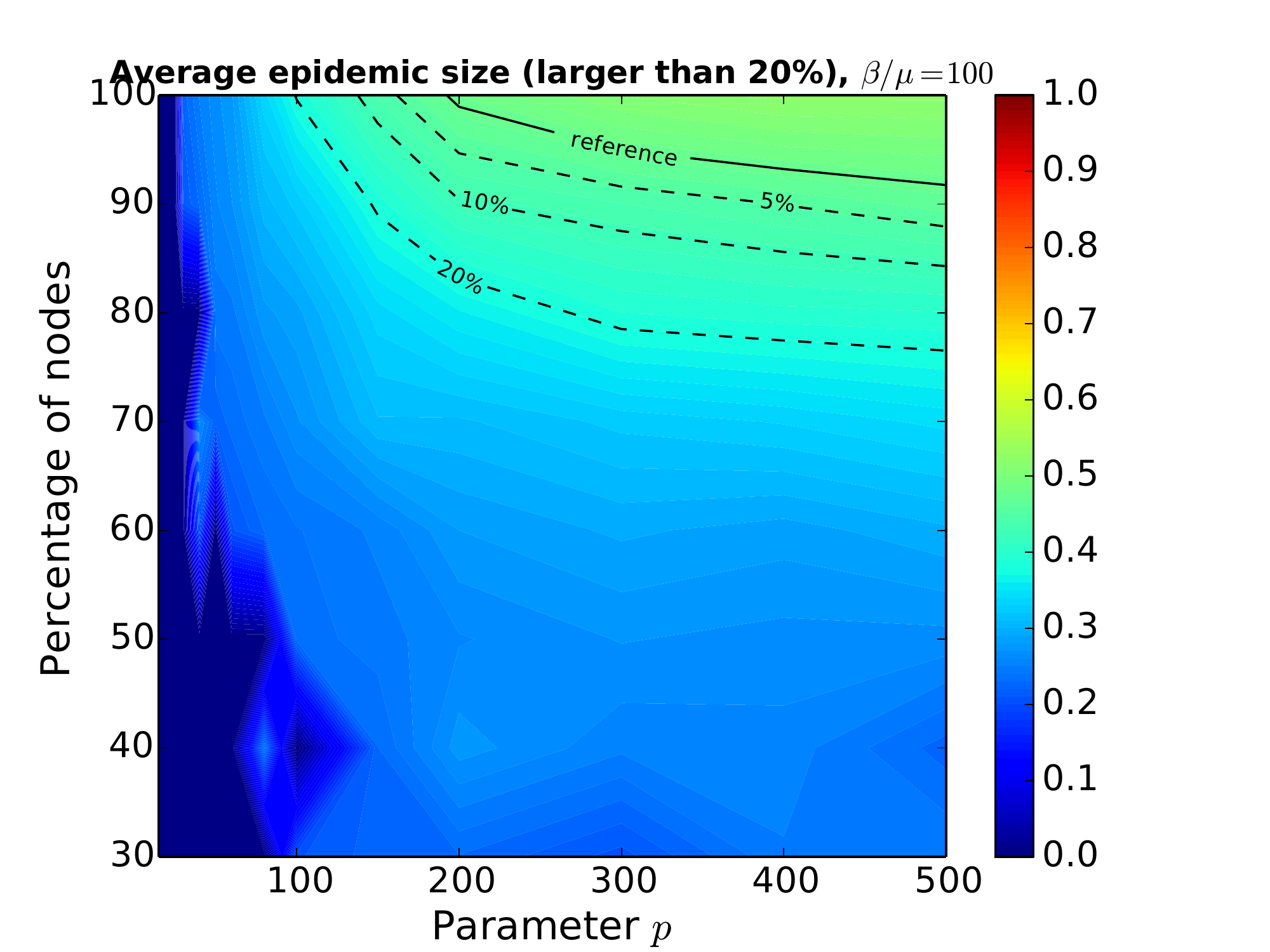}}
\subfigure[]
{\includegraphics[width=0.5\textwidth]{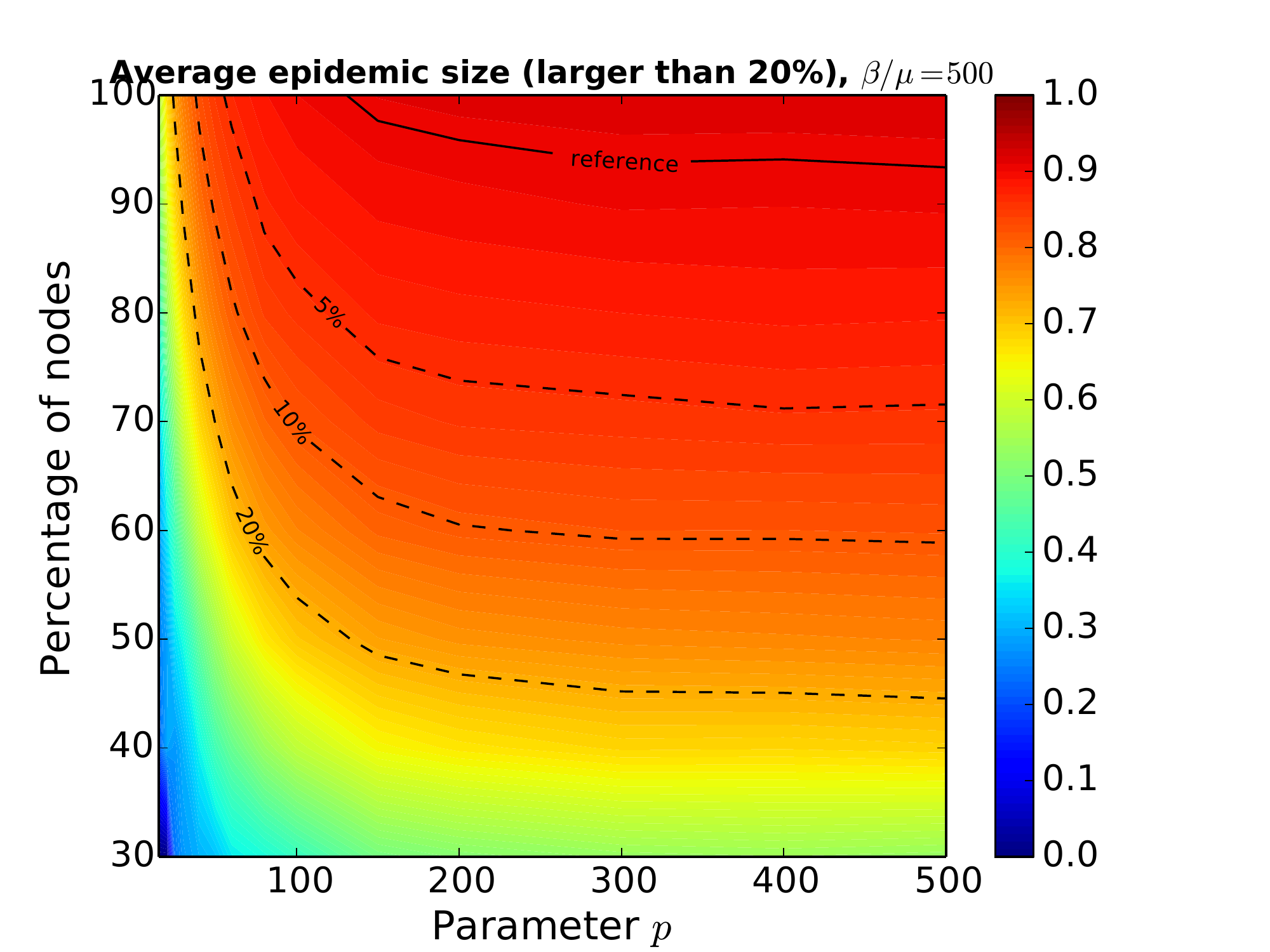}}\\
\caption{\textbf{Color maps of the average epidemic size for epidemics with size above 20\% for several
values of $\beta/\mu$ (SFHH data set).}
When no epidemics has size above 20\%, the value is zero. The three dashed lines represent the value of
average epidemic size at 5\%, 10\%, 20\% of the reference value (solid line),
which corresponds to the average epidemic size of the SIR spreading simulations performed on the contact network.\label{fig:s16}}
\end{figure}

We moreover show in Fig.s \ref{fig:s15}-\ref{fig:s16} the equivalent of Fig. 8 of the main text for these two data sets, highlighting
the combined effects of population sampling and of the absence of links with small weights in the sampled network.

\begin{table}[!ht]
\begin{tabular}{|l||l|l|l|l|}
\hline
& Number of nodes & Number of edges & Density \\
\hline
SFHH Contact network & 403 & 9565 & 0.12  \\
SFHH EGOref network & 165 & 766 & 0.06  \\
SFHH RN network & 165 & 1598 & 0.12  \\
\hline
InVS Contact network & 92 & 755 & 0.18  \\
InVS EGOref network & 37 & 88 & 0.13  \\
InVS RN network & 37 & 119 & 0.18  \\
\hline
\end{tabular}
\caption{Number of nodes and edges in the empirical and sampled data sets. The sampled networks consider the same
fraction of nodes as for the data set used in the main text.
\label{tab:s1}}
\end{table}

\end{document}